\documentclass[journal]{IEEEtran}
\usepackage{cite}
\usepackage{enumerate}
\usepackage{amsmath}
\usepackage{amsfonts}
\usepackage{mathrsfs}
\usepackage{amssymb}
\usepackage{cases}
\usepackage{stmaryrd}
\usepackage{mathrsfs}
\usepackage{bbm}
\usepackage{algorithm}
\usepackage{algorithmic} 
\usepackage{multirow}
\usepackage{graphicx}
\usepackage{float}
\usepackage{subfig}
\usepackage{bm}
\usepackage{color}
\usepackage{setspace} 
\usepackage[acronym]{glossaries}
\usepackage{booktabs}
\usepackage[colorlinks]{hyperref}
\usepackage{threeparttable}
\usepackage{multirow}
\usepackage{setspace}
\usepackage{makecell}
\usepackage{xcolor}
\usepackage{pagecolor}
\usepackage{lipsum}
\begin{document}
	
	\title{Intelligent Wiretap Code Design: Exploiting Wireless Endogenous Security via Information Theory and Deep Learning Integration}
	
	\author{Haibin Zhang,
		Xiangnan Zhou,
		Chao Wang,~\IEEEmembership{Senior Member,~IEEE,}
		Liang Jin,
		Hao Xu,\\
		Yao Sun,
		Chonghua Wang, Derrick Wing Kwan Ng,~\IEEEmembership{Fellow,~IEEE,}
		and Giuseppe Caire,~\IEEEmembership{Fellow,~IEEE}

		\thanks{H. B. Zhang, X. N. Zhou, and C. Wang are with Xidian University, Xi'an 710071, China (e-mail: hbzhang@mail.xidian.edu.cn; zhouxiangnan2021@163.com; drchaowang@126.com).}
		\thanks{L. Jin is with Information Engineering University, Zhengzhou, 450001, and also with Peng Cheng Laboratory, Shenzhen 518066, China (e-mail: liangjin@263.net).}

		\thanks{H. Xu is with the National Mobile Communications Research Laboratory, Southeast University, Nanjing, 210096, China (e-mail: hao.xu@ucl.ac.uk).}
		\thanks{Y. Sun is with the James Watt School of Engineering, University of Glasgow, Glasgow G12 8QQ, U.K. (e-mail: Yao.Sun@glasgow.ac.uk).}
		\thanks{C. H. Wang is with the China Industrial Control Systems Cyber Emergency Response Team, Beijing, 100029, China (e-mail: chonghuaw@live.com).}
		\thanks{D. W. K. Ng is with the School of Electrical Engineering and Telecommunications, University of New South Wales, Sydney, NSW 2052, Australia (e-mail: w.k.ng@unsw.edu.au).}

		\thanks{G. Caire is with the Faculty of Electrical Engineering and Computer Science at the Technical University of Berlin, 10587 Berlin, Germany (e-mail: caire@tuberlin.de).}
		\thanks{This paper was presented in part in the proceedings of the IEEE Global Communications Conference (GLOBECOM) Workshops 2024 \cite{2024wiretap}.}}

	\maketitle
	
	\begin{abstract}
		Recent advancements in wireless endogenous security have explored leveraging the inherent randomness of wireless channels to enhance communication security, providing an effective alternative to traditional encryption methods. This paper proposes a wiretap coding scheme within the semantic communication framework, which leverages discrete semantic representations compatible with conventional digital modulation to jointly enhance communication security and reliability. We investigate two eavesdropping scenarios: (i) the eavesdropper employs a maximum a posteriori (MAP) decoder, and (ii) the eavesdropper   has access to  a decoder identical to that of the legitimate receiver.
		In the first scenario, we exploit mutual information as a metric to guide the design of an optimized coding strategy, minimizing information leakage while enhancing communication reliability. In the second scenario, considering the limitations of the eavesdropper's decoding capability, we employ generalized mutual information (GMI) to characterize recoverability under the prescribed decoding rule and guide reliability-aware code optimization.
		{\color{blue}Simulation results demonstrate that the proposed approach significantly outperforms traditional separate coding schemes and   representative end-to-end methods in balancing  communication security and reliability.}
		{\color{blue}These results   demonstrate the practical value  of data-driven joint optimization in secure coding design   and provide guidance for  future secure semantic communication systems.}
	\end{abstract}
	
	
	\section{Introduction}
	\IEEEPARstart{W}{ireless} communication plays a crucial role in both military and civilian sectors, where safeguarding sensitive information against unauthorized access poses a formidable challenge \cite{a1}. Although traditional encryption techniques have been extensively researched and widely implemented \cite{a2}, they encounter several practical challenges, such as complexities in key management, computational complexity, signaling overhead, and stringent protocol design requirements \cite{a6}. Consequently, there is an increasing interest in implementing wireless endogenous security \cite{endogenous1,endogenous2}, which functions independently of upper-layer protocols, serving as a complementary or alternative approach to safeguard information security. {\color{blue}In this context, wiretap coding has   emerged as a promising approach for achieving information-theoretic security by exploiting channel randomness and the advantage  of the legitimate   link over the eavesdropping link  \cite{a6}.}
	
	The wiretap channel model, originally proposed by Wyner, provides a foundational theoretical framework for wiretap coding \cite{a21}. This model involves a transmitter (Alice), a legitimate receiver (Bob), and an eavesdropper (Eve) \cite{wyner}. Wyner demonstrated that, for memoryless channels, secure communication is achievable when the wiretap channel (from Alice to Eve) is a degraded version of the main channel (from Alice to Bob), provided the codeword length is sufficiently large. Under these conditions, Bob can decode the transmitted information with negligible error, while ensuring that Eve gains no meaningful information.  Based on Wyner's theory, researchers have   developed  a range of practical coding schemes for secure communication   by exploiting the properties of conventional  error-correcting codes. {\color{blue}Representative coding approaches include  polar wiretap coding (PWC), which leverages the polarization effect   \cite{a16}, and low-density parity-check (LDPC) codes, which exploit channel disparities and random perturbations \cite{a13}.} 
	
	However, the performance of these coding schemes, which build upon the assumption of arbitrarily long   codewords, often deviates from expectations in practical finite-length scenarios, especially for short codes. Moreover,   their inherent fixed parameters  may constrain the ability to flexibly balance reliability and security \cite{a17}, motivating alternative methods  better aligned with real-world constraints. {\color{blue}To address this, deep learning offers a unified framework by  directly embedding reliability and security objectives into the loss function   for joint optimization. Initial neural wiretap coding studies mainly adopted autoencoder-based architectures for Gaussian channels; for  instance, \cite{a17}   improved Bob's decoding  reliability by minimizing the  reconstruction  mean squared error (MSE)   while constraining Eve's information leakage  through a Taylor-series approximation of mutual information.}   {\color{blue}Since this approximation is mainly tailored to low signal-to-noise ratio (SNR) regimes, subsequent studies integrated neural mutual-information estimators \cite{a18, a19, a22} to improve applicability over  a wider range of SNR conditions.}  {\color{blue}In parallel with these end-to-end designs, a modular neural wiretap coding architecture was developed by separating neural reliability coding from hash-based secrecy processing \cite{a20}. Building on this modular design, subsequent studies extended the framework to helper-assisted Gaussian wiretap channels \cite{rana2025helper} and multi-tap Rayleigh fading wiretap channels \cite{seifert2025modular}.}

	\IEEEpubidadjcol

	{\color{blue}The aforementioned methods primarily   follow a conventional separation-based architecture, in which source coding and secure channel coding are designed independently. Although such a design is theoretically justified by  Shannon's separation theorem   in the asymptotic blocklength regime \cite{Shannon}, its asymptotic optimality does not generally extend to practical finite-blocklength transmissions. Residual channel-decoding errors can arise under non-ideal channel conditions. Without joint adaptation, such errors may propagate through the source decoding process, causing severe degradation or even complete failure of source reconstruction  \cite{Joint, JSCC}.}
	
	{\color{blue}To   circumvent these limitations, joint source-channel coding (JSCC) coordinates source representation and channel transmission within a coherent framework, moving beyond conventional separate processing. Leveraging deep learning techniques, deep joint source-channel coding (DeepJSCC) implements this co-design principle through an end-to-end neural architecture that unifies feature extraction and channel mapping, allowing transmission quality to degrade gracefully as channel conditions deteriorate \cite{JSCC}.}
	{\color{blue}Semantic communication further extends this paradigm by shifting the core criterion from exact symbol-level accuracy toward semantic fidelity and task utility, aligned with Weaver's foundational framework \cite{weaver1949recent}. This evolution reconfigures transceiver design into a joint source-channel-task optimization problem, thereby consolidating semantic feature learning, physical transmission, and downstream task execution into a cohesive framework \cite{xie2021deepSC,gunduz2023beyond}.}

	{\color{blue}This paradigm shift also reshapes the nature of secrecy threats in wiretap channels, extending the security focus beyond the recovery of source signals or bits to the inference of sensitive source attributes and task-related information. In response, secure semantic communication extends the end-to-end security mechanisms developed in secure joint source-channel coding (secure JSCC), such as adversarial training, visual protection, and privacy-constrained optimization \cite{Adversarial,secure_deepjscc_multi_eve,visual_protection_jscc,privacy_aware_jscc}, to semantic representations and downstream tasks. Subsequent  studies have extended   this line of research to security evaluation and more diverse protection mechanisms. For instance,   semantic secrecy outage probability has been proposed to characterize semantic information leakage beyond conventional bit- or signal-level measures \cite{du2022rethinking}. Multi-task learning and adversarial perturbations have also been exploited to reduce the dependence between task-relevant semantic representations and sensitive source attributes \cite{privacy_multitask_semcom}. Moreover, security-aware training,  physical-layer security   constraints, and rate-distortion-equivocation analysis have been incorporated into end-to-end semantic transmission frameworks~\cite{JSCA,DeepSSC,secure_semantic_wiretap}.}
	
	Despite recent progress, several key limitations remain. {\color{blue}First,   many existing studies optimize security against a prescribed Eve-side neural decoder \cite{Adversarial, JSCA, DeepSSC}, which may underestimate information leakage when Eve has access to a decoder matched to Eve's observation model. Second,  a gap persists between deep-learning-based secure coding and classical coset coding, as existing neural designs still lack an information-theoretic formulation that incorporates the randomized structure into coding optimization. Finally,  the reliance   of existing end-to-end schemes on continuous-valued channel representations for differentiable optimization limits their compatibility with conventional digital modulation and coding pipelines.}
	
	{\color{blue}Collectively, these representative research directions advance secure transmission along divergent dimensions. Neural wiretap coding primarily addresses bit-level physical-layer confidentiality. Secure DeepJSCC methods emphasize end-to-end source-channel co-design, learning a direct mapping from source data to channel symbols through a unified neural architecture to jointly optimize transmission and privacy protection. Semantic communication security focuses on protecting task-related information and sensitive attributes, typically through adversarial training or learning-based privacy constraints. However, systematic design methods that jointly integrate semantic representation learning and physical-layer secure coding within an information-theoretically guided framework, while remaining compatible  with conventional digital modulation   and coding pipelines, remain limited. Against this background, this paper develops an information-theoretically guided secure semantic communication framework centered on a discrete semantic interface. Through this interface, structured coset randomization is incorporated into end-to-end coding optimization, while tractable reliability and information-leakage objectives are formulated based on information-theoretic criteria.} 
	The main contributions of this work are   summarized as follows \footnote{ {\color{blue}Compared with  the conference   version, this journal article has been substantially extended  by: 1)   deriving  a new   information-leakage upper bound tailored to  the   coset coding structure, thereby providing analytical guidance for randomized neural encoding; 2) investigating a system setting in which  the   eavesdropper employs a  decoder   instance identical to that of  the legitimate receiver,  together  with a   corresponding  optimization strategy;  and 3)  further  validating the extensibility and flexibility of   the proposed  framework through  more comprehensive  experiments   and analyses, including additional evaluation metrics,  video transmission tasks,  and   reliability-security  trade-off   studies.}}:
	\begin{itemize}
		\item {\color{blue}We develop a novel secure semantic communication scheme that integrates discrete semantic representations with the randomized structure of coset coding, thereby establishing an end-to-end secure coding paradigm going beyond conventional adversarial-training-based and privacy-constrained designs. This paradigm is compatible with conventional digital modulation and facilitates practical implementation. It also offers inherent flexibility to navigate the tradeoff between semantic transmission security and reliability.}
		
		\item {\color{blue}We investigate two representative eavesdropper capability models: an optimal-decoder eavesdropper that approximates MAP decoding and an identical-decoder eavesdropper that applies the legitimate receiver's decoding metric. For these two models, tractable optimization objectives are formulated based on block-wise  mutual information and   generalized mutual information (GMI), respectively, to   jointly optimize transmission  reliability and security. More importantly, we derive a novel information-leakage upper bound specifically tailored to the  coset coding structure, thereby establishing a theoretically grounded bridge between classical coset coding and deep-learning-based secure coding.}
		
		\item {\color{blue}Extensive experimental results demonstrate that the proposed   schemes significantly outperform the considered separation-based and joint source-channel  coding (JSCC)   baselines. Under the optimal-decoder eavesdropper model, the proposed method achieves an average reliability improvement of at least 30\%  over the separation-based scheme employing conventional compression and PWC. Ablation studies that combine the proposed semantic coding architecture with PWC further show consistently improved security performance across the tested SNR range.}
	\end{itemize}
	
	The rest of the paper is organized as follows. Section~II introduces the system model and the related problem formulations. Section~III presents the coding design for the scenario where the eavesdropper is capable of performing MAP decoding. Section~IV explores the coding scheme design for the scenario where the eavesdropper and the legitimate receiver use an identical decoder instance. Extensive numerical results validating the proposed framework are provided in Section~V. Finally, Section~VI concludes the paper.
	
	Notation:   $\mathbf{I}_n$ denotes the $n \times n$  identity matrix, and   $\mathcal{CN}$ denotes the complex Gaussian  distribution. The sets of real and complex numbers are denoted by   $\mathbb{R}$ and $\mathbb{C}$, respectively. The expectation   and entropy operators are denoted by $\mathbb{E}[\cdot]$ and $H(\cdot)$, respectively. Bold uppercase letters, such as $\mathbf{X}$, $\mathbf{Y}$, and $\mathbf{Z}$, denote  random variables, whereas   the corresponding  bold lowercase letters, such as $\mathbf{x}$, $\mathbf{y}$, and $\mathbf{z}$, denote their realizations. The Kullback-Leibler  (KL) divergence between   the distributions $p(\cdot)$ and $q(\cdot)$ is defined as $\mathrm{KL}\!\left[p(\cdot)\,\|\,q(\cdot)\right] = \mathbb{E}_{p(\cdot)}\!\left[\ln p(\cdot)-\ln q(\cdot)\right]$. {\color{blue}The main variables and trainable parameters used throughout this paper are summarized in Table~\ref{tab:notation}.}
	
	\begin{table}[!t]
		\centering
		\caption{Notation.}
		\label{tab:notation}
		\scriptsize
		\renewcommand{\arraystretch}{1.18}
		\setlength{\tabcolsep}{1.2pt}
		
		\begin{tabular}{p{0.22\columnwidth}p{0.68\columnwidth}}
			\toprule
			\textbf{Symbol} & \textbf{Description} \\
			\midrule
			$\mathbf{X}$ & Source data \\
			$\mathbf{Y}$ & Task-relevant information \\
			$\mathbf{V}$ & Semantic feature  \\
			$\mathbf{Z}$ & Information block \\
			$\tilde{\mathbf{Z}}$ & Randomization block \\
			$\mathbf{S}$ & Codeword \\
			$\mathbf{Z}_{\mathrm{B}}$, $\mathbf{Z}_{\mathrm{E}}$ & Channel outputs at Bob and Eve \\
			$\mathbf{V}_{\mathrm{B}}$, $\mathbf{V}_{\mathrm{E}}$ & Recovered semantic features at Bob and Eve \\
			$\theta,\Phi$ & Parameters of the semantic encoder and decoder \\
			$\phi,\tau$ & Parameters of the channel encoder and decoder \\
			$q_\vartheta, q_\kappa$ & Information leakage estimators \\
			\bottomrule
		\end{tabular}
	\end{table}
	
	\begin{figure*}[!htb]
		\centering
		\includegraphics[width=1.0\linewidth]{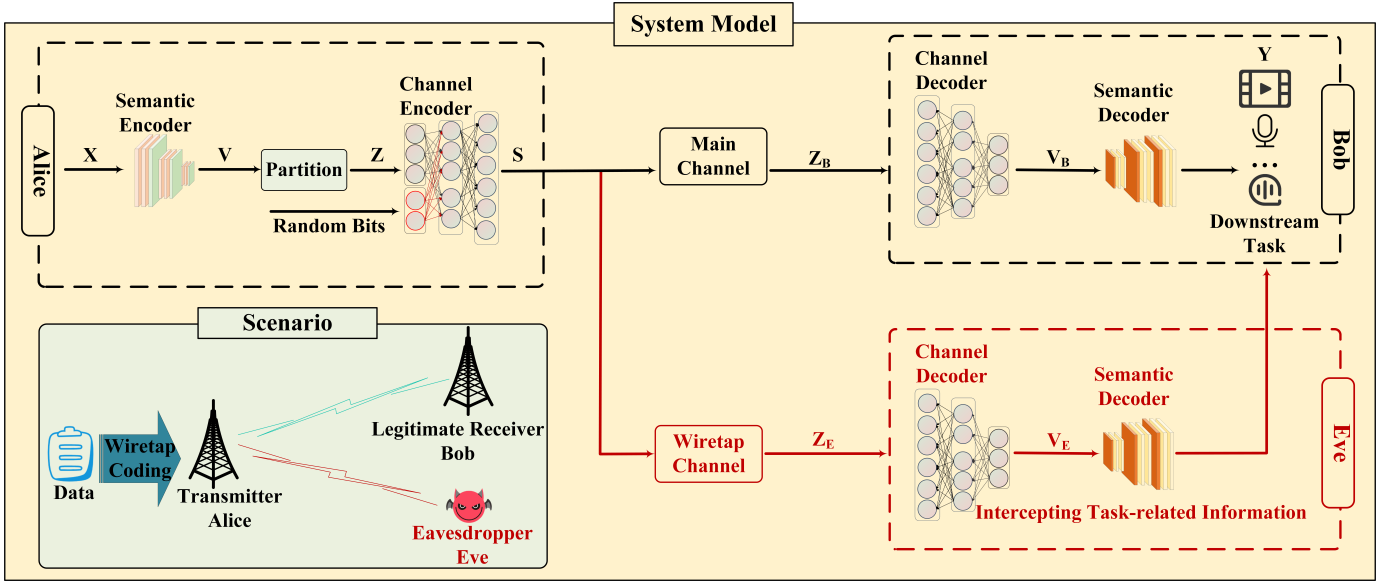}
		\caption{Illustration  of the proposed conceptual scenario and the corresponding system model.}
		\label{fig:system_model}
	\end{figure*}
	\section{Communication Scenario and System Model}
	Consider the communication scenario depicted in Fig.~\ref{fig:system_model}, where Alice serves as the transmitter, Bob is the legitimate receiver, and Eve is the eavesdropper. 
	The system's primary goal is to ensure the secure and reliable transmission of confidential information from Alice to Bob. This goal is achieved through  coding techniques that enhance decoding accuracy while mitigating the risk of interception by Eve.
	
	At the transmitter, Alice encodes the data \(\mathbf{X} \in \mathbb{R}^H \) into a semantic feature vector \(\mathbf{V} \in \{0,1\}^L \)   using  a semantic encoder \(p_\theta(\cdot) \), where \(H \) and \(L \) denote the dimensions of \(\mathbf{X} \) and \(\mathbf{V} \), respectively. Subsequently, for channel encoding, Alice employs a block coding scheme, segmenting \(\mathbf{V} \) into \(B \) blocks, each comprising \(k \) information bits.   {\color{blue}Inspired by Wyner's classic coset coding paradigm \cite{wyner}, Alice introduces structured coding randomness into the learning-based transmission to suppress information leakage. Specifically, each $k$-bit  information block $\mathbf{Z}$ is augmented with an independent, privately generated $r$-bit randomization block $\tilde{\mathbf{Z}}$ that remains unknown to both Bob and Eve.  The concatenated sequence is then mapped by the neural channel encoder $p_\phi(\cdot)$ into an $n$-bit  codeword   $\mathbf{S}$, effectively assigning each information block onto a coset containing $2^r$ candidate codewords.}
	
	{\color{blue}Following Wyner's foundational framework \cite{wyner,a21}, both the main and wiretap links  are modeled as additive white Gaussian noise (AWGN) channels\footnote{ \color{blue}The Gaussian wiretap channel model is   adopted here as an idealized setting for  its analytical   tractability  and   widespread use  in   physical-layer security.    More   general scenarios, such as  fading  or non-degraded  channels,   present distinct characteristics and may yield a non-positive secrecy capacity; they remain beyond  the   main scope of this study.}. To enable transmission over these channels, each binary codeword is modulated into a complex-valued symbol vector   using  a conventional digital modulation scheme\footnote{ \color{blue} In this work,  binary phase shift keying (BPSK)  is selected as the default modulation scheme for simplicity and alignment with prior studies. The  proposed scheme can be extended to support other conventional modulation formats, such as quadrature phase-shift keying (QPSK)  and  16-ary quadrature amplitude modulation (16-QAM). For notational simplicity, identical symbols are adopted for each codeword and its BPSK-modulated channel input vector in  derivations, except when the modulation mapping is explicitly formulated.}. The   wiretap link is assumed to be a stochastically degraded version of the main link. Specifically, the channel impairments are characterized by independent complex additive white Gaussian noise vectors $\boldsymbol{\epsilon}_\mathrm{B} \sim \mathcal{CN}(\mathbf{0}, \sigma_\mathrm{B}^2 \mathbf{I}_n)$ and $\boldsymbol{\epsilon}_\mathrm{E} \sim \mathcal{CN}(\mathbf{0}, \sigma_\mathrm{E}^2 \mathbf{I}_n)$, where the noise variances satisfy $\sigma_\mathrm{E}^2 > \sigma_\mathrm{B}^2$ to formalize the degraded eavesdropping environment.}

	{\color {blue} Upon receiving the channel output \(\mathbf{Z}_\mathrm{B} \in \mathbb{C}^n\), Bob uses a channel decoder \(q_\tau(\cdot)\) to retrieve the information blocks, which are subsequently aggregated into the reconstructed semantic feature vector \(\mathbf{V}_\mathrm{B} \in \{0,1\}^L\). Finally, a semantic decoder \(q_\Phi(\cdot)\) is employed at Bob to obtain an estimate of the task-relevant information \(\mathbf{Y}\). Here, \(\mathbf{Y}\) denotes the task-relevant target associated with the source sample \(\mathbf{X}\); depending on the downstream task, it may correspond to a class label, a semantic attribute, or the target content for reconstruction in image/video transmission. Simultaneously, Eve observes the stochastically degraded signal \(\mathbf{Z}_\mathrm{E} \in \mathbb{C}^n\). To formalize the secrecy breach, Eve processes this observation via its dedicated channel and semantic processing chain, reconstructing the feature vector \(\mathbf{V}_\mathrm{E} \in \{0,1\}^L\) to construct its estimate of \(\mathbf{Y}\).}
	
	The  specific optimization objectives for achieving communication reliability and security   are detailed in the subsequent sections. To ground  this analysis,   the functional  dependency structure among   the system variables is explicitly characterized by  the following Markov   chains:
	\begin{align}
		\mathbf{Y} \rightarrow \mathbf{X}  \rightarrow \mathbf{Z} \rightarrow \mathbf{S} \rightarrow \mathbf{Z}_\mathrm{B} \rightarrow \mathbf{V}_\mathrm{B}, \\
		\mathbf{Y} \rightarrow \mathbf{X}  \rightarrow \mathbf{Z} \rightarrow \mathbf{S} \rightarrow \mathbf{Z}_\mathrm{E} \rightarrow \mathbf{V}_\mathrm{E}.
	\end{align} 
	This generative structure   implicitly governs  the joint distribution over all variables involved in both training and evaluation, with respect to which all  expectations, mutual information quantities, and variational objectives throughout this paper are defined. For notational  simplicity, the   intermediate variable \(\mathbf{V}\) is omitted henceforth without altering the core information flow. {\color{blue}
		Furthermore, the source-task joint distribution
		\(p(\mathbf{x},\mathbf{y})\) is assumed to be given, serving as the base from which other induced joint distributions are derived. For example, the joint distribution can be expressed as
		\begin{align}
			p(\mathbf{z},\mathbf{z}_{\mathrm E})
			&=
			\int \int \sum_{\mathbf{s}}
			p(\mathbf{y},\mathbf{x},\mathbf{z},\mathbf{s},\mathbf{z}_{\mathrm E})
			\,\mathrm{d}\mathbf{x}\,\mathrm{d}\mathbf{y}
			\notag\\
			&=
			\int \int \sum_{\mathbf{s}}
			p(\mathbf{y},\mathbf{x})
			p(\mathbf{z}\mid\mathbf{x})
			p(\mathbf{s}\mid\mathbf{z})
			p(\mathbf{z}_{\mathrm E}\mid\mathbf{s})
			\,\mathrm{d}\mathbf{x}\,\mathrm{d}\mathbf{y}.
		\end{align}
	}

	\section{Coding Schemes for the Eavesdropper Exploiting the Optimal Decoder}
	{\color{blue}This section considers the joint reliability-and-security optimization problem under an optimal-decoding eavesdropper model. The eavesdropper is assumed to know the coding structure and the statistical characteristics of the wiretap channel, allowing Eve to employ a MAP decoder matched to the observation distribution. Knowledge of the coding structure is consistent with Kerckhoffs' principle, whereas access to the statistical characteristics of the wiretap channel represents an additional strong assumption regarding Eve's capabilities.}

	\subsection{Reliability Optimization}

	To ensure reliable data transmission, it is essential that the codeword block \(\mathbf{S} \), after being transmitted through the main channel, can be accurately decoded by Bob to recover the original information block \(\mathbf{Z} \). To this end, the channel encoder at Alice  is designed to preserve the fidelity of the transmitted information, thereby enabling accurate reconstruction of \(\mathbf{Z} \) at Bob's receiver. {\color{blue}From an information-theoretic   viewpoint, this reliability  objective can be   formalized  by maximizing the mutual information   $I(\mathbf{Z}; \mathbf{Z}_\mathrm{B})$, which quantifies the  reduction in the uncertainty of   $\mathbf{Z}$ provided by the legitimate observation $\mathbf{Z}_\mathrm{B}$.} Formally, the mutual information   $I(\mathbf{Z}; \mathbf{Z}_\mathrm{B})$ is given by
	\begin{align}
		I(\mathbf{Z}; \mathbf{Z}_\mathrm{B}) 
		= \sum_{\mathbf{z}}\int p(\mathbf{z}, \mathbf{z}_\mathrm{B}) \ln \frac{p(\mathbf{z}, \mathbf{z}_\mathrm{B})}{p(\mathbf{z}) p(\mathbf{z}_\mathrm{B})} \,\mathrm{d}\mathbf{z}_\mathrm{B}.
	\end{align} 
	
	However, while ensuring symbol-level transmission reliability is crucial, 
	practical constraints such as nonideal channel conditions and finite codeword 
	lengths may still lead to unavoidable decoding errors, thereby degrading 
	downstream task performance.
	{\color{blue}
	Therefore, beyond conventional symbol-level reliability metrics, it is 
	imperative that the recovered semantic feature 
	\(\mathbf{V}_{\mathrm B}\) at Bob preserves the task-critical information 
	required to infer \(\mathbf{Y}\), even in the presence of channel-induced 
	distortions.
	This requirement can be addressed by maximizing the mutual information 
	\(I(\mathbf{Y};\mathbf{V}_{\mathrm B})\) between the task target 
	\(\mathbf{Y}\) and the recovered semantic feature 
	\(\mathbf{V}_{\mathrm B}\). To this end, the joint reliability objective is 
	formulated as
	\begin{align}
		\underset{\theta,\phi,\tau,\Phi}{\max}\quad
		I(\mathbf{Y};\mathbf{V}_{\mathrm B})
		+\alpha_1 I(\mathbf{Z};\mathbf{Z}_{\mathrm B}),
		\label{eq:joint_reliability_objective}
	\end{align}
	where \(\theta\), \(\phi\), \(\tau\), and \(\Phi\) denote the trainable 
	parameters of the respective neural network modules, and \(\alpha_1>0\) is 
	a weighting coefficient that balances the two reliability metrics.
	}
	
		{\color{blue}However, the mutual information terms in 
		\eqref{eq:joint_reliability_objective} are generally intractable, as the 
		involved posterior distributions lack closed-form expressions~\cite{a17}.
		We thus derive a tractable lower bound for the joint reliability objective. 
		Since \(H(\mathbf{Y})\) is independent of all trainable parameters, maximizing 
		\eqref{eq:joint_reliability_objective} is equivalent to maximizing the shifted 
		objective
	\begin{align}
		\bar{J}_{\mathrm R}
		&=
		I(\mathbf{Y};\mathbf{V}_{\mathrm B})
		+\alpha_1 I(\mathbf{Z};\mathbf{Z}_{\mathrm B})
		-H(\mathbf{Y})
		\notag\\
		&=
		-H(\mathbf{Y}\mid\mathbf{V}_{\mathrm B})
		+\alpha_1
		\{
		\underbrace{H(\mathbf{Z})}_{\geq 0}
		-H(\mathbf{Z}\mid\mathbf{Z}_{\mathrm B})
		\}
		\notag\\
		&\geq
		-H(\mathbf{Y}\mid\mathbf{V}_{\mathrm B})
		-\alpha_1 H(\mathbf{Z}\mid\mathbf{Z}_{\mathrm B}).
		\label{eq:reliability_entropy_bound}
	\end{align}
	Here, the inequality follows from \(H(\mathbf{Z})\geq 0\). 
	To further derive computable expressions, we introduce the neural-network-parameterized variational distributions
		$q_\Phi(\mathbf{y}|\mathbf{v}_{\mathrm B})$ and
		$q_\tau(\mathbf{z}|\mathbf{z}_{\mathrm B})$. The corresponding conditional entropies can be expressed as
	\begin{align}
		&H(\mathbf{Y}|\mathbf{V}_{\mathrm B}) \notag\\
		&=
		-\mathbb{E}_{p(\mathbf{y},\mathbf{v}_{\mathrm B})}
		\left[\ln q_\Phi(\mathbf{y}|\mathbf{v}_{\mathrm B})
		\right]-
		\mathbb{E}_{p(\mathbf{v}_{\mathrm B})}
		\left[\mathrm{KL}\left(p(\mathbf{y}|\mathbf{v}_{\mathrm B})
		\middle\|
		q_\Phi(\mathbf{y}|\mathbf{v}_{\mathrm B})
		\right)
		\right],
		\label{eq:semantic_variational_decomposition}\\
		&H(\mathbf{Z}|\mathbf{Z}_{\mathrm B}) \notag\\
		&=
		-\mathbb{E}_{p(\mathbf{z},\mathbf{z}_{\mathrm B})}
		\left[\ln q_\tau(\mathbf{z}|\mathbf{z}_{\mathrm B})
		\right]-
		\mathbb{E}_{p(\mathbf{z}_{\mathrm B})}
		\left[\mathrm{KL}\left(p(\mathbf{z}|\mathbf{z}_{\mathrm B})
		\middle\|
		q_\tau(\mathbf{z}|\mathbf{z}_{\mathrm B})
		\right)
		\right].
		\label{eq:channel_variational_decomposition}
	\end{align}
	Combining \eqref{eq:reliability_entropy_bound} with
	\eqref{eq:semantic_variational_decomposition} and
	\eqref{eq:channel_variational_decomposition}, together with the non-negativity of the conditional KL-divergence terms, gives the tractable lower bound
	\begin{align}
		\bar{J}_{\mathrm R} 
		\geq
		\mathbb{E}_{p(\mathbf{y},\mathbf{v}_{\mathrm B})}
		\left[\ln q_\Phi(\mathbf{y}|\mathbf{v}_{\mathrm B})
		\right] +
		\alpha_1
		\mathbb{E}_{p(\mathbf{z},\mathbf{z}_{\mathrm B})}
		\left[\ln q_\tau(\mathbf{z}|\mathbf{z}_{\mathrm B})
		\right].
		\label{eq:tractable_reliability_bound}
	\end{align}
	This formulation replaces the intractable true posteriors
	\(p(\mathbf{y} | \mathbf{v}_{\mathrm{B}})\) and
	\(p(\mathbf{z} | \mathbf{z}_{\mathrm{B}})\) with the
	neural-network-parameterized variational approximations
	\(q_{\Phi}(\mathbf{y} | \mathbf{v}_{\mathrm{B}})\) and
	\(q_{\tau}(\mathbf{z} | \mathbf{z}_{\mathrm{B}})\), respectively,
	yielding a tractable reliability objective that can be estimated from
	training samples. The corresponding approximation gaps are quantified
		by the conditional KL-divergence terms in
		\eqref{eq:semantic_variational_decomposition} and
		\eqref{eq:channel_variational_decomposition}. Accordingly, maximizing
		the  variational   log-likelihood terms reduces these KL-divergence gaps
		and improves the accuracy of the posterior approximations.}

	\subsection{Security Optimization and Information Leakage Estimation}
	{\color{blue}Under the considered model, Eve's inference capability is fundamentally bounded by the statistical dependence between the information block $\mathbf{Z}$ and Eve's received signal $\mathbf{Z}_\mathrm{E}$.
    To quantify secrecy risks, we employ mutual information $I(\mathbf{Z};\mathbf{Z}_\mathrm{E})$ to characterize information leakage. This quantity evaluates the average reduction in uncertainty of $\mathbf{Z}$ conditioned on $\mathbf{Z}_\mathrm{E}$, and constitutes a decoder-independent information-theoretic security criterion.
	Analogous to the reliability analysis, directly optimizing $I(\mathbf{Z};\mathbf{Z}_\mathrm{E})$ is computationally intractable owing to inaccessible true posteriors. To derive optimization-compatible security metrics, we utilize a Jensen-type leakage estimator and a coset-structure-based leakage upper bound within the training objective.
	Accordingly, these two leakage terms are incorporated into the optimization objective to constrain information leakage from complementary perspectives.}
	\subsubsection{Jensen-Type Variational Estimator for Information Leakage}
	
	{\color{blue}
		Neural network-parameterized estimators for mutual information bounds have been extensively investigated in recent studies, with the Contrastive Log-ratio Upper Bound (CLUB) serving as a prominent approach~\cite{Club}.
		Specifically, by applying Jensen's inequality  to the logarithmic term involving
		\(\mathbb{E}_{p(\mathbf{z})}[p(\mathbf{z}_\mathrm{E}|\mathbf{z})]\), an upper
		bound on information leakage can be derived and expressed as follows:
		\begin{align}
			I(\mathbf{Z}; \mathbf{Z}_\mathrm{E})
			&=\mathbb{E}_{p(\mathbf{z}, \mathbf{z}_\mathrm{E})}
			\left[\ln p(\mathbf{z}_\mathrm{E}|\mathbf{z}) \right]
			- \mathbb{E}_{p(\mathbf{z}_\mathrm{E})}
			\left[\ln p(\mathbf{z}_\mathrm{E}) \right] \notag \\ 
			&=\mathbb{E}_{p(\mathbf{z}, \mathbf{z}_\mathrm{E})}
			\left[\ln p(\mathbf{z}_\mathrm{E}|\mathbf{z}) \right]
			- \mathbb{E}_{p(\mathbf{z}_\mathrm{E})}
			\left[\ln \mathbb{E}_{p(\mathbf{z})}
			\left[p(\mathbf{z}_\mathrm{E}|\mathbf{z})\right] \right] \notag \\ 
			&\leq
			\mathbb{E}_{p(\mathbf{z}, \mathbf{z}_\mathrm{E})}
			\left[\ln p(\mathbf{z}_\mathrm{E}|\mathbf{z}) \right]
			- \mathbb{E}_{p(\mathbf{z})}\mathbb{E}_{p(\mathbf{z}_\mathrm{E})}
			\left[\ln p(\mathbf{z}_\mathrm{E}|\mathbf{z}) \right].
		\end{align} 
		Replacing  the true conditional distribution   $p(\mathbf{z}_{\mathrm{E}}|\mathbf{z})$ with a parameterized variational distribution $q_{\vartheta}(\mathbf{z}_{\mathrm{E}}|\mathbf{z})$ yields the following trainable variational leakage estimator:
		\begin{align}
			U_b \approx
			\mathbb{E}_{p(\mathbf{z}, \mathbf{z}_\mathrm{E})}
			\left[\ln q_\vartheta(\mathbf{z}_\mathrm{E}|\mathbf{z})\right]
			- \mathbb{E}_{p(\mathbf{z})}\,\mathbb{E}_{p(\mathbf{z}_\mathrm{E})}
			\left[\ln q_\vartheta(\mathbf{z}_\mathrm{E}|\mathbf{z})\right].
		\end{align}

		The discrepancy between the resulting trainable estimator and the true information leakage arises from two sources: the introduced Jensen relaxation and the variational approximation of  the true conditional distribution   by $q_{\vartheta}(\mathbf{z}_{\mathrm{E}}|\mathbf{z})$  \cite{Club}. To   minimize  the   variational gap, we reduce the KL divergence from the true conditional distribution to $q_{\vartheta}(\mathbf{z}_{\mathrm{E}}|\mathbf{z})$, leading to the following objective:

		\begin{align}
			\underset{\vartheta}{\min}\quad
			&\mathbb{E}_{p(\mathbf{z})}
			\left[\mathrm{KL}
			\left(p(\mathbf{z}_\mathrm{E}|\mathbf{z})
			\| q_\vartheta(\mathbf{z}_\mathrm{E}|\mathbf{z})
			\right)
			\right] \notag \\
			=
			\underset{\vartheta}{\min}\quad
			&\underbrace{\mathbb{E}_{p(\mathbf{z},\mathbf{z}_\mathrm{E})}
				\left[\ln p(\mathbf{z}_\mathrm{E}|\mathbf{z})
				\right]}_{\text{const. w.r.t. } \vartheta}
			-
			\mathbb{E}_{p(\mathbf{z},\mathbf{z}_\mathrm{E})}
			\left[\ln q_\vartheta(\mathbf{z}_\mathrm{E}|\mathbf{z})
			\right] \notag \\
			=
			\underset{\vartheta}{\min}\quad
			&-\mathbb{E}_{p(\mathbf{z},\mathbf{z}_\mathrm{E})}
			\left[\ln q_\vartheta(\mathbf{z}_\mathrm{E}|\mathbf{z})
			\right].\label{var}
		\end{align}

		\subsubsection{Upper Bound on Information Leakage Based on Coset Coding Structure}

		Leveraging the Markov relation established in the system model and the coset coding structure, the information leakage can be upper-bounded  as follows:
		\begin{align}
			I(\mathbf{Z};\mathbf{Z}_{\mathrm{E}})
			&= I(\mathbf{Z},\tilde{\mathbf{Z}};\mathbf{Z}_{\mathrm{E}})
			-I(\tilde{\mathbf{Z}};\mathbf{Z}_{\mathrm{E}}|\mathbf{Z})
			\label{eq:leakage_start}\\
			&= I(\mathbf{S};\mathbf{Z}_{\mathrm{E}})
			-I(\tilde{\mathbf{Z}};\mathbf{Z}_{\mathrm{E}}|\mathbf{Z})
			\label{eq:leakage_step2}\\
			&= I(\mathbf{S};\mathbf{Z}_{\mathrm{E}})
			-H(\tilde{\mathbf{Z}}|\mathbf{Z})
			+H(\tilde{\mathbf{Z}}|\mathbf{Z},\mathbf{Z}_{\mathrm{E}})
			\label{eq:leakage_step3}\\
			&= I(\mathbf{S};\mathbf{Z}_{\mathrm{E}})
			-H(\tilde{\mathbf{Z}})
			+H(\tilde{\mathbf{Z}}|\mathbf{Z},\mathbf{Z}_{\mathrm{E}})
			\label{eq:leakage_step4}\\
			&\leq \mathcal{C}_{\mathrm{E}}
			-H(\tilde{\mathbf{Z}})
			-\mathbb{E}_{p(\mathbf{z},\tilde{\mathbf{z}},\mathbf{z}_{\mathrm{E}})}
			\left[\ln q_{\kappa}
			(\tilde{\mathbf{z}}|\mathbf{z},\mathbf{z}_{\mathrm{E}})
			\right]
			\notag\\
			&\quad
			-\mathbb{E}_{p(\mathbf{z},\mathbf{z}_{\mathrm{E}})}
			\{\underbrace{\mathrm{KL}\!\left(p(\tilde{\mathbf{z}}|\mathbf{z},\mathbf{z}_{\mathrm{E}})
				\middle\|
				q_{\kappa}(\tilde{\mathbf{z}}|\mathbf{z},\mathbf{z}_{\mathrm{E}})
				\right)}_{\geq 0}
			\}
			\label{eq:leakage_step5}\\
			&\leq \mathcal{C}_{\mathrm{E}}
			-\underbrace{H(\tilde{\mathbf{Z}})}_{\mathrm{constant}}
			-\mathbb{E}_{p(\mathbf{z},\tilde{\mathbf{z}},\mathbf{z}_{\mathrm{E}})}
			\left[\ln q_{\kappa}
			(\tilde{\mathbf{z}}|\mathbf{z},\mathbf{z}_{\mathrm{E}})
			\right],
			\label{eq:leakage_end}
		\end{align} 
		where   \(\mathcal{C}_\mathrm{E} = \underset{p(\mathbf{s})\in\mathcal{S}}{\max} \ I(\mathbf{S}; \mathbf{Z}_\mathrm{E})\). Here, \(\mathcal{S}\) denotes the set of all probability distributions on \(\{0,1\}^{n}\) whose support has cardinality at most \(2^{k+r}\). Thus, maximizing \(I(\mathbf{S}; \mathbf{Z}_{\mathrm{E}})\) over \(\mathcal{S}\) yields the maximum information leakage achievable by any input distribution respecting this coding constraint. The term \(q_\kappa(\tilde{\mathbf{z}} | \mathbf{z}, \mathbf{z}_\mathrm{E})\) represents a variational approximation of \(p(\tilde{\mathbf{z}} | \mathbf{z}, \mathbf{z}_\mathrm{E})\).
	}
	{\color{blue}The main steps in the above derivation are clarified as follows:}
	\begin{itemize}
		\item 
		{\color{blue}The derivation from \eqref{eq:leakage_start} to \eqref{eq:leakage_step2} follows from the identity $I(\mathbf{Z},\tilde{\mathbf{Z}};\mathbf{Z}_{\mathrm{E}})=I(\mathbf{S};\mathbf{Z}_{\mathrm{E}})$, which relies on the deterministic channel encoding and the Markov chain $(\mathbf{Z},\tilde{\mathbf{Z}})\to\mathbf{S}\to\mathbf{Z}_{\mathrm{E}}$. Specifically, since $\mathbf{S}$ is a deterministic function of $(\mathbf{Z},\tilde{\mathbf{Z}})$, we have $H(\mathbf{Z}_{\mathrm{E}}|\mathbf{Z},\tilde{\mathbf{Z}})=H(\mathbf{Z}_{\mathrm{E}}|\mathbf{Z},\tilde{\mathbf{Z}},\mathbf{S})$. Moreover, the Markov chain implies that $\mathbf{Z}_{\mathrm{E}}$ is conditionally independent of the pre-encoding variables $(\mathbf{Z},\tilde{\mathbf{Z}})$ given the transmitted codeword $\mathbf{S}$, yielding $H(\mathbf{Z}_{\mathrm{E}}|\mathbf{Z},\tilde{\mathbf{Z}},\mathbf{S})=H(\mathbf{Z}_{\mathrm{E}}|\mathbf{S})$. Therefore, $H(\mathbf{Z}_{\mathrm{E}}|\mathbf{Z},\tilde{\mathbf{Z}})=H(\mathbf{Z}_{\mathrm{E}}|\mathbf{S})$, and hence $I(\mathbf{Z},\tilde{\mathbf{Z}};\mathbf{Z}_{\mathrm{E}})=I(\mathbf{S};\mathbf{Z}_{\mathrm{E}})$.}
		\item {\color{blue}The transition from   \eqref{eq:leakage_step3} to \eqref{eq:leakage_step4} follows from the independence between \(\tilde{\mathbf{Z}}\) and \(\mathbf{Z}\), which gives \(H(\tilde{\mathbf{Z}}|\mathbf{Z})=H(\tilde{\mathbf{Z}})\). Equation \eqref{eq:leakage_step4} illustrates the role of randomized encoding. A portion of the information transmitted via Eve's channel is consumed to eliminate the uncertainty stemming from randomization, thereby diminishing the extractable information regarding \(\mathbf{Z}\) contained in the channel observations.}
		\item {\color{blue}The upper bound in \eqref{eq:leakage_step5} is obtained by bounding the mutual information and conditional entropy terms in \eqref{eq:leakage_step4} separately. To begin with, the mutual information \(I(\mathbf{S};\mathbf{Z}_\mathrm{E})\) arising from the learned encoder is upper-bounded by Eve's channel capacity \(\mathcal{C}_\mathrm{E}\) over \(\mathcal{S}\). The conditional entropy \(H(\tilde{\mathbf{Z}}|\mathbf{Z},\mathbf{Z}_\mathrm{E})\) is further decomposed via the variational distribution \(q_{\kappa}(\tilde{\mathbf{z}}|\mathbf{z},\mathbf{z}_\mathrm{E})\), and the nonnegativity of the corresponding conditional KL divergence delivers a tractable variational bound.}
	\end{itemize}

	{\color{blue}The tightness of the derived bound can be evaluated from two perspectives: the capacity relaxation gap between the mutual information induced by the practical codeword distribution and the maximum mutual information over the family of admissible input distributions, and the conditional KL divergence between $q_{\kappa}(\tilde{\mathbf{z}}|\mathbf{z},\mathbf{z}_\mathrm{E})$ and the true posterior $p(\tilde{\mathbf{z}}|\mathbf{z},\mathbf{z}_\mathrm{E})$. Following \eqref{var}, $q_{\kappa}$ is optimized by minimizing this KL divergence, which narrows the latter gap:
		\begin{align}
			\underset{\kappa}{\min}\quad
			&\mathbb{E}_{p(\mathbf{z},\mathbf{z}_\mathrm{E})}
			\left[\mathrm{KL}\!\left(p(\tilde{\mathbf{z}}|\mathbf{z},\mathbf{z}_\mathrm{E})
			\| q_\kappa(\tilde{\mathbf{z}}|\mathbf{z},\mathbf{z}_\mathrm{E})
			\right)
			\right] \notag \\
			=
			\underset{\kappa}{\min}\quad
			&\underbrace{\mathbb{E}_{p(\mathbf{z},\tilde{\mathbf{z}},\mathbf{z}_\mathrm{E})}
				\left[\ln p(\tilde{\mathbf{z}}|\mathbf{z},\mathbf{z}_\mathrm{E})
				\right]}_{\text{const. w.r.t. } \kappa}
			-
			\mathbb{E}_{p(\mathbf{z},\tilde{\mathbf{z}},\mathbf{z}_\mathrm{E})}
			\left[\ln q_\kappa(\tilde{\mathbf{z}}|\mathbf{z},\mathbf{z}_\mathrm{E})
			\right] \notag \\
			=
			\underset{\kappa}{\min}\quad
			&-\mathbb{E}_{p(\mathbf{z},\tilde{\mathbf{z}},\mathbf{z}_\mathrm{E})}
			\left[\ln q_\kappa(\tilde{\mathbf{z}}|\mathbf{z},\mathbf{z}_\mathrm{E})
			\right].
	\end{align} }

	\subsection{Joint Optimization for Reliability and Security}

	\begin{figure*}[!ht]
		\begin{align}
			\underset{\theta, \phi, \tau, \Phi}{\max}\quad
			&
			\mathbb{E}_{p(\mathbf{y}, \mathbf{v}_{\mathrm{B}})}
			\left[\ln q_\Phi(\mathbf{y}|\mathbf{v}_{\mathrm{B}})
			\right]
			+
			\alpha_1
			\mathbb{E}_{p(\mathbf{z}, \mathbf{z}_{\mathrm{B}})}
			\left[\ln q_\tau(\mathbf{z}|\mathbf{z}_{\mathrm{B}})
			\right]
			-
			\alpha_2
			\Big\{\mathbb{E}_{p(\mathbf{z}, \mathbf{z}_{\mathrm{E}})}
			\left[\ln q_\vartheta(\mathbf{z}_{\mathrm{E}}|\mathbf{z})
			\right]
			-
			\mathbb{E}_{p(\mathbf{z})}
			\mathbb{E}_{p(\mathbf{z}_{\mathrm{E}})}
			\left[\ln q_\vartheta(\mathbf{z}_{\mathrm{E}}|\mathbf{z})
			\right]
			\Big\}
			\notag \\
			&
			+
			\alpha_3
			\mathbb{E}_{p(\mathbf{z}, \tilde{\mathbf{z}}, \mathbf{z}_{\mathrm{E}})}
			\left[\ln q_\kappa(\tilde{\mathbf{z}}|\mathbf{z},\mathbf{z}_{\mathrm{E}})
			\right].
			\label{eq:joint_objective}
		\end{align} 
	\end{figure*}
	{\color{blue}To enhance the reception reliability at Bob and suppress information leakage to Eve, we formulate the joint optimization problem in \eqref{eq:joint_objective} at the top of the next page. The weighting coefficients \(\alpha_2>0\) and \(\alpha_3>0\) dictate  the trade-off between reliability and   secrecy. Since the two leakage-related terms depend on their dedicated estimator networks, the corresponding variational approximations evolve alongside encoder parameter updates, which necessitates periodic re-optimization of all leakage estimators following the update rule in \eqref{eq:variational_update}. We therefore adopt an alternating optimization scheme: the encoder is optimized via \eqref{eq:joint_objective}, while the estimator networks are updated using \eqref{eq:variational_update}. This iterative alternating workflow is elaborated in lines~9--14 of Algorithm~\ref{alg:alg1}.}
	\begin{align}
		\min_{\vartheta, \kappa} \quad -\mathbb{E}_{p(\mathbf{z}, \tilde{\mathbf{z}}, \mathbf{z}_{\mathrm{E}})} \left[\ln q_{\vartheta}(\mathbf{z}_{\mathrm{E}}|\mathbf{z}) + \ln q_{\kappa}(\tilde{\mathbf{z}}|\mathbf{z},\mathbf{z}_{\mathrm{E}}) \right].
		\label{eq:variational_update}
	\end{align}

	To enable end-to-end training, the expectations in the optimization objectives are approximated through  Monte Carlo sampling  \footnote{ \color{blue} Estimating $\mathbb{E}_{p(\mathbf{z})}\mathbb{E}_{p(\mathbf{z}_\mathrm{E})}[\ln q_\vartheta(\mathbf{z}_\mathrm{E}|\mathbf{z})]$ requires sampling independently from both marginal distributions. To approximate this expectation in practice, we draw paired instances from the joint distribution $p(\mathbf{z}, \mathbf{z}_\mathrm{E})$ and construct pseudo-independent pairs by applying a randomized permutation index $t=t(i,b)$ to $\mathbf{z}_\mathrm{E}$. This batch-shuffling strategy, widely adopted in contrastive learning, offers a practical alternative to direct marginal sampling without incurring additional sampling overhead~\cite{a22, shuffle}.}. {\color{blue}For notational convenience, we define the block-level empirical averaging operator as
		\begin{align}
			\widehat{\mathbb{E}}
			\left[g_{i,b,u,j}
			\right]
			\triangleq
			\frac{1}{N}
			\sum_{i=1}^{N}
			\frac{1}{B}
			\sum_{b=1}^{B}
			\frac{1}{U}
			\sum_{u=1}^{U}
			\frac{1}{M}
			\sum_{j=1}^{M}
			g_{i,b,u,j}.
			\label{eq:empirical_average}
		\end{align}
		Here, \(N\) denotes the number of training pairs, \(U\) denotes the number of sampled randomization blocks for each information block, and \(M\) denotes the number of channel realizations sampled for each codeword. Based on this operator, the joint reliability-and-security objective in \eqref{eq:joint_objective} and the leakage estimator objective in \eqref{eq:variational_update} are reformulated as the loss functions \(L_1\) and \(L_2\), given in \eqref{eq:L1} at the top of the next page and \eqref{eq:L2}, respectively.} 
	\begin{align}
		L_2
		=
		\widehat{\mathbb{E}}
		\Big[&
		-\ln q_{\vartheta}
		\left(\mathbf{z}_{\mathrm{E}}^{(i,b,u,j)}
		|
		\mathbf{z}^{(i,b)}
		\right)
		\notag \\
		&
		-
		\ln q_{\kappa}
		\left(\tilde{\mathbf{z}}^{(i,b,u)}
		|
		\mathbf{z}^{(i,b)},
		\mathbf{z}_{\mathrm{E}}^{(i,b,u,j)}
		\right)
		\Big].
		\label{eq:L2}
	\end{align}	
	The corresponding sampling procedure  is detailed in lines   3--8 of Algorithm~\ref{alg:alg1}.
	\begin{figure*}[!ht]
		\vspace{-5mm}
		\begin{align}
			L_1
			=
			\widehat{\mathbb{E}}
			\Bigg[&
			-\ln q_{\Phi}
			\left(\mathbf{y}^{(i)}
			|
			\mathbf{v}_{\mathrm{B}}^{(i,u,j)}
			\right)
			-
			\alpha_1
			\ln q_{\tau}
			\left(\mathbf{z}^{(i,b)}
			|
			\mathbf{z}_{\mathrm{B}}^{(i,b,u,j)}
			\right)
			+
			\alpha_2
			\Big\{\ln q_{\vartheta}
			\left(\mathbf{z}_{\mathrm{E}}^{(i,b,u,j)}
			|
			\mathbf{z}^{(i,b)}
			\right)
			-
			\ln q_{\vartheta}
			\left(\mathbf{z}_{\mathrm{E}}^{(t(i,b),u,j)}
			|
			\mathbf{z}^{(i,b)}
			\right)
			\Big\}
			\notag \\
			&
			-
			\alpha_3
			\ln q_{\kappa}
			\left(\tilde{\mathbf{z}}^{(i,b,u)}
			|
			\mathbf{z}^{(i,b)},
			\mathbf{z}_{\mathrm{E}}^{(i,b,u,j)}
			\right)
			\Bigg].
			\label{eq:L1}
		\end{align}
		\hrule
		\vspace{-5mm}
	\end{figure*}

	\subsection{Complexity Analysis and Training Strategy}
	{\color{blue}The computational complexity of the proposed scheme is analyzed for the offline training and online inference phases. Let $L_{\rm se}$, $L_{\rm sd}$, $L_{\rm c}$, and $L_{\rm v}$ denote the total layer counts of the semantic encoder, semantic decoder, neural channel encoder-decoder pair, and two estimators $q_{\vartheta}$  and   $q_{\kappa}$, respectively, and let $d_{\rm se}$, $d_{\rm sd}$,  $d_{\rm c}$, and $d_{\rm v}$ denote their maximum hidden dimensions. The corresponding network architectures are detailed in Section~V. For an \(L\)-layer network with maximum input or output dimension \(d\), the per-sample forward-pass complexity is \(O(Ld^{2})\).
		
		During offline training, the coding networks are updated $T_{1}$ times, and each coding-network update is accompanied by $T_{2}$ estimator updates. For $N$ training samples, \(U\) sampled randomization blocks per information block, and \(M\) channel realizations per codeword, the  overall   training complexity is
		\begin{align}
			O\left(NT_{1}
			\left[L_{\rm se}d_{\rm se}^{2}
			+
			U M L_{\rm sd}d_{\rm sd}^{2}
			+
			B U M
			\left(L_{\rm c}d_{\rm c}^{2}
			+
			T_{2}L_{\rm v}d_{\rm v}^{2}
			\right)
			\right]
			\right).
		\end{align}
		During online inference, no iterative optimization is required, and each source sample only undergoes a forward pass through the coding networks. The resulting online complexity is
		\begin{align}
			O\left(L_{\rm se}d_{\rm se}^{2}
			+
			L_{\rm sd}d_{\rm sd}^{2}
			+
			B L_{\rm c}d_{\rm c}^{2}
			\right).
		\end{align}
	}

	To  enable gradient-based training through the  multiple quantization operations in the encoding process, we   employ  the straight-through estimator (STE)~\cite{STE}. {\color{blue}However, the  alternating  optimization of the   coding networks and leakage estimators may result in unstable updates of the leakage-related objective. This instability can be further aggravated by the gradient bias introduced by the STE.} {\color{blue}To address this  issue, we adopt a three-stage training strategy: first, the semantic   parameters \(\theta\) and \(\Phi\)  are trained in a noise-free environment; second, the semantic module is fixed, and the channel   parameters \(\phi\) and \(\tau\)  are trained based on samples drawn from the   block-level  probability distribution produced by the semantic module; finally, after these initialization stages, the third-stage training follows Algorithm~\ref{alg:alg1}, where the leakage estimators and coding networks are alternately updated.}

	\begin{algorithm}[htb]
		\setstretch{1.07}
		{\color{blue}
		\caption{Sampling and Training}\label{alg:alg1}
		\begin{algorithmic}[1]
			\STATE {\bf Input:} Training iterations \(T_1,T_2\), sampled randomization count \(U\), channel-realization count \(M\), and noise variances \(\sigma_\mathrm{B}^2\), \(\sigma_\mathrm{E}^2\).
			\FOR{$e = 1$ to $T_1$}
			\STATE Sample \(N\) pairs   \(\{(\mathbf{x}^{(i)}, \mathbf{y}^{(i)})\}_{i=1}^N \sim p(\mathbf{x}, \mathbf{y})\).
			\STATE Compute semantic features \(\{\mathbf{v}^{(i)}\}_{i=1}^N\) and partition them into \(N \times B\) blocks \(\{\mathbf{z}^{(i,b)}\}_{i=1,b=1}^{N,B}\).
			\STATE   For each information block, sample \(U\) independent randomization blocks
			$\{\tilde{\mathbf{z}}^{(i,b,u)}\}_{i=1,b=1,u=1}^{N,B,U}$
			and generate the corresponding codewords
			$\{\mathbf{s}^{(i,b,u)}\}_{i=1,b=1,u=1}^{N,B,U}$.
			\STATE Generate \(M\) noise samples \(\boldsymbol{\epsilon}_\mathrm{B}\) and \(\boldsymbol{\epsilon}_\mathrm{E}\) for the transmission of each codeword   \(\mathbf{s}^{(i,b,u)}\).
			
			\STATE Compute received signals   \(\{\mathbf{z}_\mathrm{B}^{(i,b,u,j)}\}_{i,b,u,j}^{N,B,U,M}\) and \(\{\mathbf{z}_\mathrm{E}^{(i,b,u,j)}\}_{i,b,u,j}^{N,B,U,M}\).
			\STATE Aggregate decoded blocks and generate \(\{\mathbf{v}_\mathrm{B}^{(i,u,j)}\}_{i,u,j}^{N,U,M}\).
			\FOR{$\ell = 1$ to $T_2$}
			\STATE Compute   $L_2$ with $\theta$, $\phi$, $\tau$, and $\Phi$ fixed.
			\STATE Update   $\vartheta$ and $\kappa$  via gradient descent.
			\ENDFOR
			\STATE Compute   $L_1$ with
			$\vartheta$ and $\kappa$ fixed.
			\STATE Update   $\theta$, $\phi$, $\tau$, and $\Phi$  via gradient descent.
			\ENDFOR
		\end{algorithmic}
		}
	\end{algorithm}

	\section{Coding Scheme   Under an Identical-Decoder  Eavesdropper}
	In this section, we consider an identical-decoder eavesdropper scenario. {\color{blue}For communication terminals employing proprietary or closed-source implementations, Eve may have limited  access to the   transmitter-side encoding mechanism. Nevertheless, the receiver-side decoding function remains vulnerable to acquisition via device capture, firmware extraction,  or reverse engineering~\cite{costin2014firmware}, which may enable Eve to execute a similar decoding pipeline over the wiretap channel.}

	\subsection{Security Optimization}
	{\color{blue} Consistent with the formulation presented in the previous section, characterizing information leakage in this setting requires evaluating the extent to which Eve can reliably infer the transmitted message from the corresponding channel observations. Mutual information provides a standard decoder-independent measure of statistical leakage, while its operational relevance to recoverability is most direct when decoding is unconstrained and matched to the wiretap channel. Under the identical-decoder model, however, Eve is restricted to the decoding rule designed for Bob, which may be mismatched to the wiretap-channel statistics. Consequently, mutual information can provide a somewhat conservative characterization of recoverability in this setting.
		
	This observation motivates the use of a decoder-aware security criterion tailored to mismatched decoding~\cite{b20,b21}. To make the dependence on the prescribed decoding rule explicit, we examine Eve's decision rule. Upon receiving the wiretap-channel output
		\(\mathbf{z}_{\mathrm E}\), the decoder evaluates the decoding metric
		\(\ln f_\tau(\mathbf{z}_{\mathrm E}\mid\mathbf{z}_i)\)
		for each candidate information block \(\mathbf{z}_i\) and selects
		\(\mathbf{z}_i\) if its metric is strictly larger than those of all other candidates:
		\begin{align}
			\ln f_\tau(\mathbf{z}_{\mathrm E}\mid\mathbf{z}_i)
			>
			\ln f_\tau(\mathbf{z}_{\mathrm E}\mid\mathbf{z}_j),
			\quad
			\forall j\neq i,
			\quad
			i,j\in\{1,2,\dots,2^k\}.
		\end{align}
		When the prescribed metric is mismatched to the wiretap channel, the induced posterior distribution generally differs from the true channel posterior, thereby limiting the reliable recovery of the transmitted information.	}
	
	Consequently,   we employ GMI to  characterize recoverability under the mismatched decoding metric.
	Specifically, we adopt the following unit-exponent GMI expression\footnote{The GMI expression used herein follows the unit-exponent form of Fischer's mismatched-decoding formulation \cite{b21} and is adopted consistently throughout the analysis and end-to-end optimization.}:
	\begin{align}
		\mathrm{GMI}(\mathbf{Z}; \mathbf{Z}_\mathrm{E}) &= \sum_{\mathbf{z}}\int p(\mathbf{z},\mathbf{z}_\mathrm{E}) \ln \left(\frac{f_\tau(\mathbf{z}_\mathrm{E} | \mathbf{z})}{\sum_{\mathbf{z}'} p(\mathbf{z}') f_\tau(\mathbf{z}_\mathrm{E} | \mathbf{z}')} \right) \,\mathrm{d}\mathbf{z}_\mathrm{E} \notag \\
		&= \sum_{\mathbf{z}}\int p(\mathbf{z},\mathbf{z}_\mathrm{E}) \ln \frac{q_\tau(\mathbf{z}|\mathbf{z}_\mathrm{E})}{p(\mathbf{z})} \,\mathrm{d}\mathbf{z}_\mathrm{E},
	\end{align} 
	where
	\begin{align}
		q_\tau(\mathbf{z}|\mathbf{z}_\mathrm{E}) = \frac{p(\mathbf{z}) f_\tau(\mathbf{z}_\mathrm{E} | \mathbf{z})}{\sum_{\mathbf{z}'} p(\mathbf{z}') f_\tau(\mathbf{z}_\mathrm{E} | \mathbf{z}')}
	\end{align}
	denotes the posterior distribution induced by the mismatched decoding metric. This expression characterizes recoverability under the prescribed decoding rule.
	
	Building on the previous analysis, we propose a protection strategy aimed at jointly reducing Eve's recoverability at both the symbol and semantic levels.
	At the symbol level, we minimize   \(\mathrm{GMI}(\mathbf{Z}; \mathbf{Z}_\mathrm{E}) \)  to weaken the eavesdropper's ability to recover the original transmitted information block \(\mathbf{Z} \) from the intercepted signal through the prescribed channel decoder. A lower value of   \(\mathrm{GMI}(\mathbf{Z}; \mathbf{Z}_\mathrm{E}) \) indicates degraded recoverability under the prescribed channel decoder, thereby limiting the information recoverable by the eavesdropper at the symbol level.
	At the semantic level, we further minimize   \(\mathrm{GMI}(\mathbf{Y}; \mathbf{V}_\mathrm{E}) \)  to suppress the eavesdropper's ability to infer task-related information from the semantic representation \(\mathbf{V}_\mathrm{E} \) under the prescribed semantic decoder. A lower value of   \(\mathrm{GMI}(\mathbf{Y}; \mathbf{V}_\mathrm{E}) \) indicates degraded task-related recoverability under the prescribed semantic decoder, thereby enhancing semantic security and reducing the risk of task-relevant information recovery.
	
	These two optimization objectives are jointly integrated into a unified loss function to enable coordinated suppression of symbol-level and semantic-level recoverability under the identical-decoder eavesdropper model. The overall optimization problem is formulated as:
	\begin{align}
		\underset{\theta, \phi, \tau, \Phi}{\min} \;
		\sum_{\mathbf{z}}\int p(\mathbf{z},\mathbf{z}_\mathrm{E}) \ln \frac{q_\tau(\mathbf{z}|\mathbf{z}_\mathrm{E})}{p(\mathbf{z})} \,\mathrm{d}\mathbf{z}_\mathrm{E} \\ \notag +
		\int p(\mathbf{y}, \mathbf{v}_\mathrm{E}) \ln \frac{q_\Phi(\mathbf{y}|\mathbf{v}_\mathrm{E})}{p(\mathbf{y})} \,\mathrm{d}\mathbf{y} \,\mathrm{d}\mathbf{v}_\mathrm{E}.
	\end{align}
	By minimizing this function, the decoder's ability to extract meaningful information from intercepted signals is   constrained, thereby improving the system's resistance to  eavesdropping and enhancing transmission security.
	
	\subsection{Reliability Optimization}
	 {\color{blue} In the previous subsection, we established a GMI-guided formulation for characterizing recoverability under mismatched decoding conditions. Under the current setup, however, involving the decoder parameters in security optimization induces a shift away from the metric matched to the main channel, thereby inevitably introducing decoder mismatch at Bob and subsequently compromising the effectiveness of the mutual information  formulation for reliability optimization.} To address this issue, we reformulate the reliability optimization problem by maximizing   \(\mathrm{GMI}(\mathbf{Z}; \mathbf{Z}_\mathrm{B}) \) and \(\mathrm{GMI}(\mathbf{Y}; \mathbf{V}_\mathrm{B}) \), which better capture the reliability performance under mismatched decoding conditions. The optimization problem is thus given as follows:
	\begin{align}
		\underset{\theta, \phi, \tau, \Phi}{\max} &\sum_{\mathbf{z}}\int p(\mathbf{z}, \mathbf{z}_\mathrm{B}) \ln \frac{q_\tau(\mathbf{z}|\mathbf{z}_\mathrm{B})}{p(\mathbf{z})} \,\mathrm{d}\mathbf{z}_\mathrm{B} \notag \\
		&\ \ \ \ + \int p(\mathbf{y}, \mathbf{v}_\mathrm{B}) \ln \frac{q_\Phi(\mathbf{y}|\mathbf{v}_\mathrm{B})}{p(\mathbf{y})} \,\mathrm{d}\mathbf{y} \,\mathrm{d}\mathbf{v}_\mathrm{B}.
	\end{align}
	
	\begin{figure*}[!ht]
		\begin{align}
			\underset{\theta, \phi, \tau, \Phi}{\max} \ 
			&\mathbb{E}_{p(\mathbf{y}, \mathbf{v}_\mathrm{B})}[\ln q_\Phi(\mathbf{y} | \mathbf{v}_\mathrm{B})] + \beta_1\mathbb{E}_{p(\mathbf{z}, \mathbf{z}_\mathrm{B})}[\ln q_\tau(\mathbf{z} | \mathbf{z}_\mathrm{B})] - \beta_2\mathbb{E}_{p(\mathbf{y}, \mathbf{v}_\mathrm{E})}[\ln q_\Phi(\mathbf{y} | \mathbf{v}_\mathrm{E})] - \beta_3\mathbb{E}_{p(\mathbf{z}, \mathbf{z}_\mathrm{E})}[\ln q_\tau(\mathbf{z} | \mathbf{z}_\mathrm{E})]. \label{eq:samejoint_objective}
		\end{align}
		\vspace{-7mm}
	\end{figure*}
	
	\begin{figure*}[!ht]
		\begin{align}
			L_3=
			\widehat{\mathbb{E}}
			\Bigg[-\ln q_\Phi(\mathbf{y}^{(i)} | \mathbf{v}_\mathrm{B}^{(i,u,j)})
			+ \beta_2 \ln q_\Phi(\mathbf{y}^{(i)} | \mathbf{v}_\mathrm{E}^{(i,u,j)})
			-\beta_1\ln q_\tau(\mathbf{z}^{(i,b)} | \mathbf{z}_\mathrm{B}^{(i,b,u,j)})
			+ \beta_3 \ln q_\tau(\mathbf{z}^{(i,b)} | \mathbf{z}_\mathrm{E}^{(i,b,u,j)})
			\Bigg]. \label{eq:L3}
		\end{align}
		\hrule
		\vspace{-3mm}
	\end{figure*}
	\subsection{Joint Optimization}		
	{
		\color{blue}
		To jointly account for transmission reliability at Bob and information security against Eve, the reliability and secrecy objectives derived above are integrated into a unified optimization framework. The resulting joint objective is given in \eqref{eq:samejoint_objective} at the top of the next page, where the hyperparameters $\beta_1>\beta_3>0$ and $\beta_2>0$ control the trade-off between transmission reliability and information security. Proceeding along the lines of Section~III, omitting the constant \(H(\mathbf{Y})\) and leveraging the non-negativity of \(H(\mathbf{Z})\) yields an optimizable lower-bound surrogate.
	}
	
	{\color{blue}
		The expectations in the joint objective are approximated by Monte Carlo averaging over the training samples, sampled randomization blocks, and channel realizations, resulting in the loss function \(L_3\) in \eqref{eq:L3} at the top of the next page. Since no auxiliary leakage estimator is required in this formulation, offline training involves only the coding and decoding networks. The resulting offline training complexity is
		\begin{align}
			O\left(
			NT_1
			\left[
			L_{\rm se}d_{\rm se}^{2}
			+
			U M L_{\rm sd}d_{\rm sd}^{2}
			+
			B U M L_{\rm c}d_{\rm c}^{2}
			\right]
			\right).
		\end{align}
		The per-sample online inference complexity is
		\begin{align}
			O\left(
			L_{\rm se}d_{\rm se}^{2}
			+
			L_{\rm sd}d_{\rm sd}^{2}
			+
			B L_{\rm c}d_{\rm c}^{2}
			\right).
		\end{align}
	}

	\section{Simulation Results}
	In this section, we evaluate the reliability and security performance of the 
	proposed scheme on  image and video transmission tasks.   The experimental 
	framework is implemented using  Python 3.9 and  PyTorch, with  an NVIDIA 
	A800 GPU   used for the computations. The underlying communication system 
	employs BPSK modulation over an  AWGN channel. {\color{blue}Tailored to semantic communications, we adopt  the mean squared error (MSE), classification accuracy, and DINOv2 feature cosine similarity (FCS) to assess semantic fidelity and semantic-leakage proxies. Notably, the ImageNet classifier and DINOv2 feature extractor~\cite{oquab2023dinov2} are employed as pretrained evaluators whose weights have been publicly released by third parties and widely adopted by the research community\footnote{The DINOv2 ViT-S/14 (\url{https://github.com/facebookresearch/dinov2}) and TorchVision ResNet-50 (\url{https://pytorch.org/vision/stable/models.html}) models are adopted for evaluation.}. Let $\psi_{\mathrm{D}}(\cdot)$ denote the representation mapping of the DINOv2 feature extractor; the FCS between the source sample $\mathbf{x}$ and its recovered counterpart $\widehat{\mathbf{x}}$ is computed as
		\begin{equation}
			\mathrm{FCS}(\mathbf{x}, \widehat{\mathbf{x}}) = \frac{\psi_{\mathrm{D}}(\mathbf{x})^{\mathrm{T}}\psi_{\mathrm{D}}(\widehat{\mathbf{x}})}{\|\psi_{\mathrm{D}}(\mathbf{x})\|_2 \|\psi_{\mathrm{D}}(\widehat{\mathbf{x}})\|_2}.
		\end{equation}
		Furthermore, to mitigate the randomness of channel transmission, the performance of each 
		sample is averaged over 100 independent channel realizations.}
	
	{\color{blue}Consistent with   established deep learning-based coding methodologies~\cite{a17,a20}, 
		we   focus on the short-blocklength regime to characterize 
		reliability and security of the proposed scheme. This setup is 
		relevant to low-latency wireless applications, such as the Internet of Things 
		(IoT) and short-packet communications, which inherently necessitate compact 
		codeword dimensions. The triplet $(g, r, n)$ is introduced to compactly represent the encoding parameters, where the aggregate block length $g = k+r$ couples the $k$-bit information block with an $r$-bit random  block.}
	\subsection{Dataset}
	We   employ  standard datasets for  the  image and video transmission tasks. {\color{blue}For image transmission,   the model is trained on a 14-class  subset of ImageNet~\cite{ImageNet} containing  14,560 images   with a resolution of (256   $\times$  256) pixels. It is evaluated  on the Kodak dataset, which contains  24 high-resolution images   of (512   $\times$  768) pixels, and an ImageNet test subset comprising  3,640 images   from the same 14 classes.} For video transmission, we   employ  the Moving MNIST dataset~\cite{Moving}, which consists of  10,000 sequences of 20 frames, with 8,000  sequences used  for training and 2,000 for testing.
	\begin{figure*}[!t]
		\centering
		\includegraphics[width=1.0\linewidth]{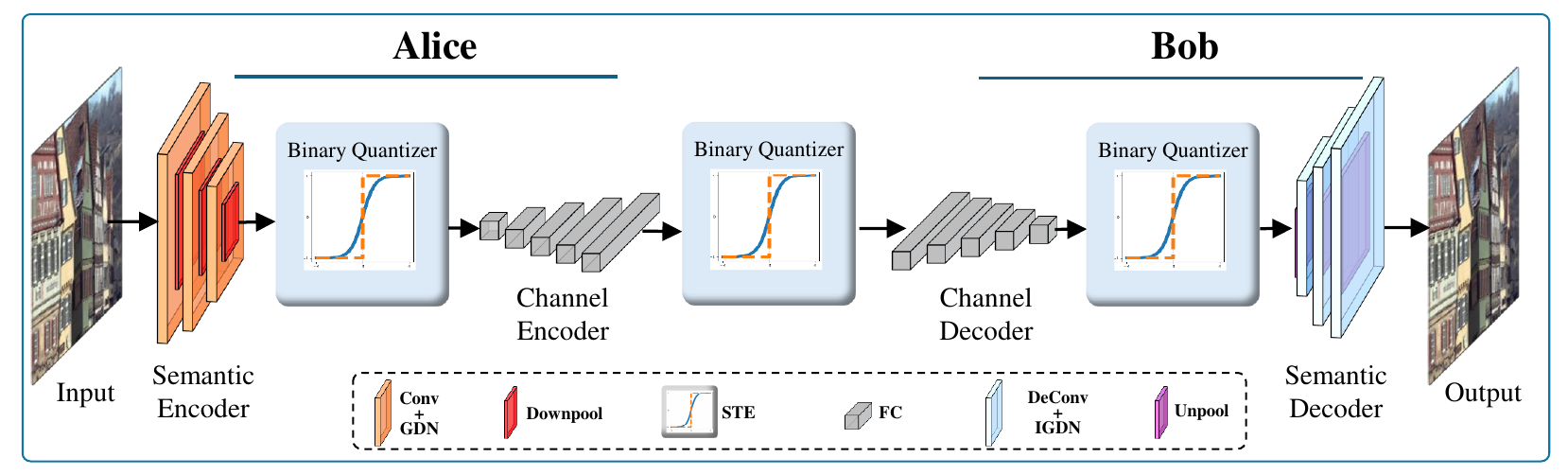}
		\caption{Model   layers.}
		\label{fig:model_layers}
	\end{figure*}
	\subsection{Model Structure}
	This work evaluates   performance across both  image and video transmission tasks. {\color{blue}Inspired by   foundational frameworks \cite{GDN, DVC, Video}, we design a compact and computationally efficient neural   architecture tailored to  the proposed information-theoretic   objective. This structural simplification avoids unnecessary computational overhead while satisfying the system performance constraints. For clarity, Fig.~\ref{fig:model_layers} illustrates  the encoding and decoding modules for the image transmission scenario, whereas  the video processing pipeline   is omitted to keep the illustration concise.}
	
	At the transmitter,   the input source (image or video frame) undergoes sequential operations 
	comprising  semantic encoding, binary quantization, and channel encoding to   yield the transmitted 
	channel symbols. For the image modality, the semantic encoder   utilizes  convolutional layers, Generalized Divisive Normalization (GDN) \cite{GDN}, and downsampling  operations  to extract compact spatial features. For video   sequences, the semantic representation comprises reference-frame  features, inter-frame motion   features, and prediction residual features.

	The resulting  continuous features are   subsequently discretized via  a binary quantizer 
	implemented through a cascaded  Tanh and Sign function. {\color{blue}To   bypass  the non-differentiability 
		of the Sign   operator  during backpropagation,  the STE \cite{STE} 
		is integrated, which preserves the   deterministic Sign operation  in the forward pass while 
		approximating   the backward gradient with an identity mapping to ensure  end-to-end 
		differentiability.} Finally,  a five-layer fully connected channel encoder  processes the 
	quantized bit blocks  to generate the   codewords.
	{\color{blue}Since codeword generation is inherently handled by the trained encoder, explicit codebook distribution and maintenance are not required. The size of the codeword set is primarily determined by the coding parameters, particularly $g$ and $n$, while the neural-network capacity dictates how effectively this mapping is represented.}

	At the receiver, Bob   applies the corresponding decoder to perform the downstream task. {\color{blue}For security evaluation, we   define the two eavesdropper models introduced above as follows: (i) a decoder   optimized independently under each operational SNR to approximate the MAP decision rule, and (ii)   an identical instance of Bob's decoder.}
	
	{\color{blue}The proposed model contains 38M trainable parameters. For input images of size \(256\times256\), the measured median per-sample network inference latency on an NVIDIA A800 GPU is approximately \(27\,\mathrm{ms}\).}

	\subsection{Baseline Schemes and Settings}
	In this section, we select four representative benchmark schemes for comparison:
	
	I. JPEG/H.264 + PWC: This scheme   uses  JPEG for image source coding and H.264 for video source coding, combined with PWC for channel coding. It represents a widely   used  method of separate encoding in practical applications. 
	
	II. {\color{blue}GDN + PWC:   Serving as an ablation baseline  for our proposed method, this architecture retains the deep semantic feature extraction network but  trains the semantic encoder  independently  in a noise-free environment,   subsequently cascading it with PWC-based channel coding.}
	
	III.   Conventional End-to-end Baseline \cite{JSCC}.
	{\color{blue}This autoencoder-driven JSCC framework directly maps source feature representations to continuous real-valued channel symbols, acting as a classic  end-to-end   benchmark.}
	
	IV. Adversarial   End-to-end Baseline \cite{Adversarial}.
	{\color{blue}Built upon an end-to-end architecture, this baseline incorporates adversarial training to constrain the eavesdropper's ability to recover useful information throughout the training phase, and it represents a mainstream secure communication benchmark.}
	
	To ensure  a  fair comparison, all  benchmark and proposed  schemes are evaluated under compression rate settings   that reflect equivalent  transmission resource constraints. The compression rate is defined as the number of  transmitted  channel symbols per source symbol, where a source symbol   corresponds  to a single pixel in an image or video frame.   For network optimization, all experiments employ  the Adam optimizer   to ensure consistent convergence. Schemes~III and IV   utilize  a learning rate of   $10^{-4}$, whereas the proposed scheme is trained with  a lower learning rate of   $10^{-5}$. 
	
	{\color{blue} Under the considered system model, Alice has no access to accurate channel statistics of Eve and adopts a static design in which SNR operating points for Bob and Eve are predefined.  These operating points are fixed and will not be adapted to accommodate drifts in channel statistics. The prescribed SNRs act as design hyperparameters and are not required to precisely match Eve's true average SNR. Unless otherwise specified, the training SNRs of Bob and Eve are set to 6~dB and 0~dB, respectively. This configuration establishes an operating point where Bob experiences a moderate SNR, whereas Eve suffers from relatively poor channel conditions throughout model optimization. The trained Alice-Bob coding network remains fixed during evaluation across different channel SNRs. Under the optimal Eve decoder setup, Eve's decoder is separately optimized for each tested SNR.}

	\begin{figure}[!t]
		\centering
		\includegraphics[width=0.85\linewidth]{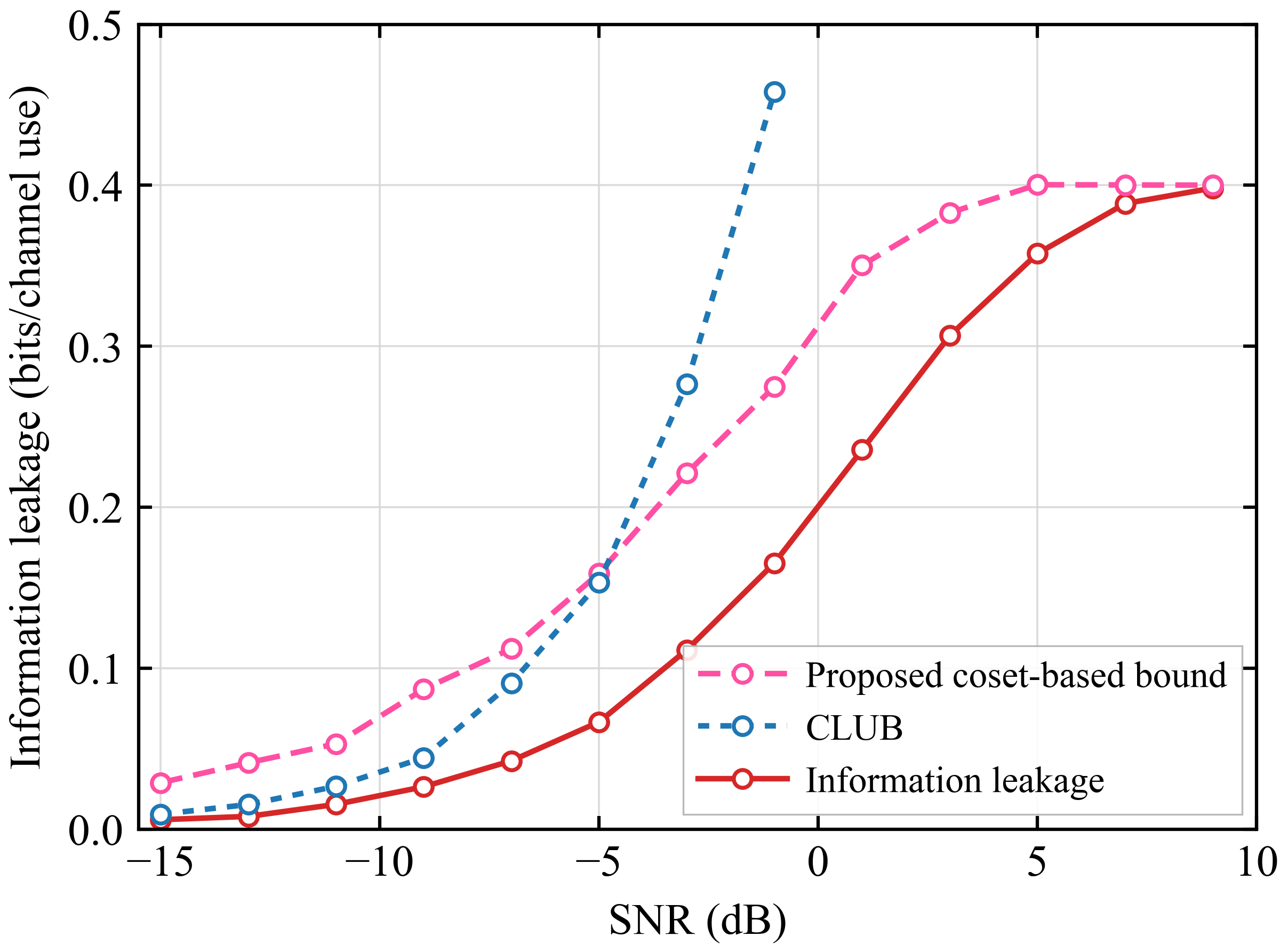}
		\caption{{\color{blue}Comparison between the benchmark information leakage, the CLUB-based estimator, and the coset-structure-based upper bound at different SNRs with \((g,r,n)=(3,1,5)\).}}
		\label{fig:leakage_bound_validation}
	\end{figure}
	\begin{figure*}[!htb]
		\centering
		\includegraphics[width=1.0\linewidth]{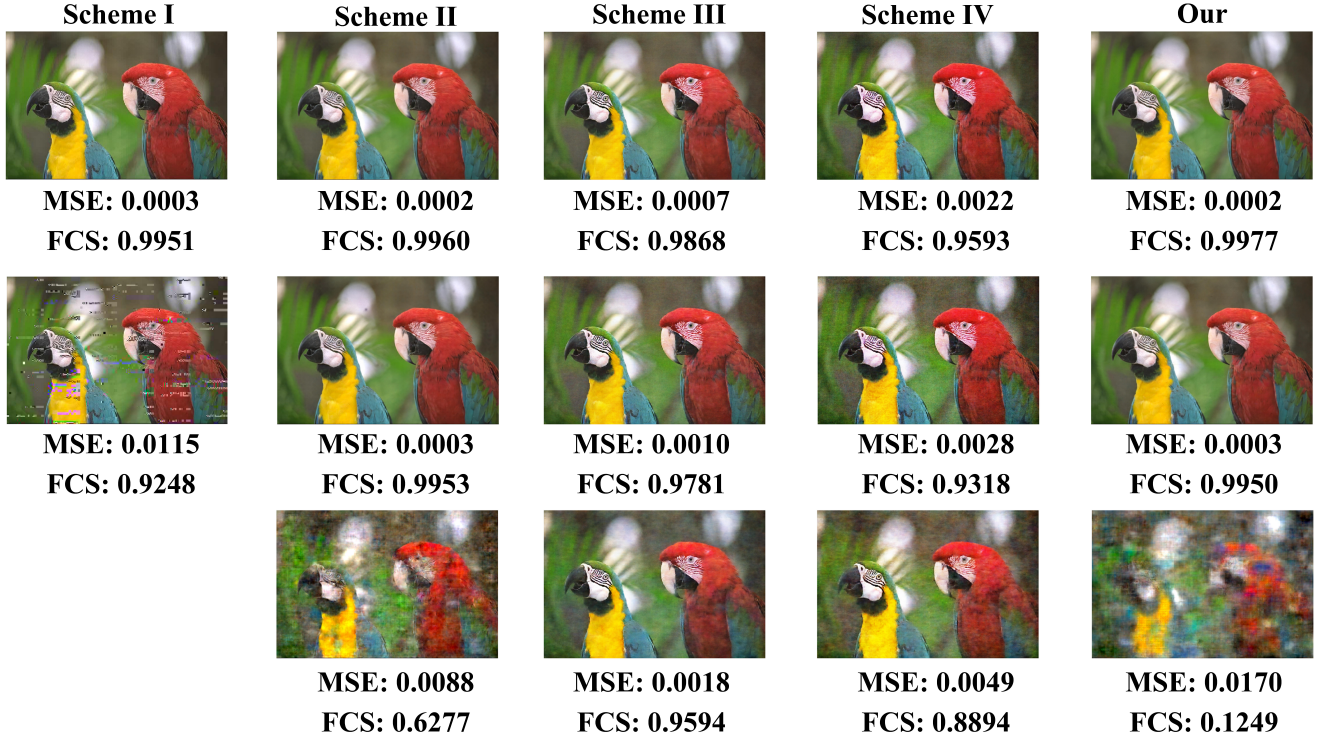}
		\caption{{\color{blue}Visual comparison  on the Kodak dataset.   From  top to bottom, the rows display the images recovered  by:  the legitimate receiver  (Bob)  at 10~dB, the  legitimate receiver  (Bob)  at 5~dB, and the  eavesdropper  (Eve)  at -3~dB. The  hyperparameters of our  scheme are   configured as $\alpha_1 = 1$,   $\alpha_2 = 20$,  and  $\alpha_3 = 4$, with  the   encoding parameters  set to   $(g, r, n) = (9, 3, 64)$.}}
		\label{fig:kodak_visual_comparison}
	\end{figure*}
	
	\subsection{Performance Evaluation with Optimal Decoder for Eve}
	In this subsection, Eve's decoder is optimized   pointwise across all tested SNRs to 
	execute MAP decoding.

	\begin{figure*}[!htb]
		\centering
		\subfloat[]{%
			\includegraphics[width=0.45\linewidth]{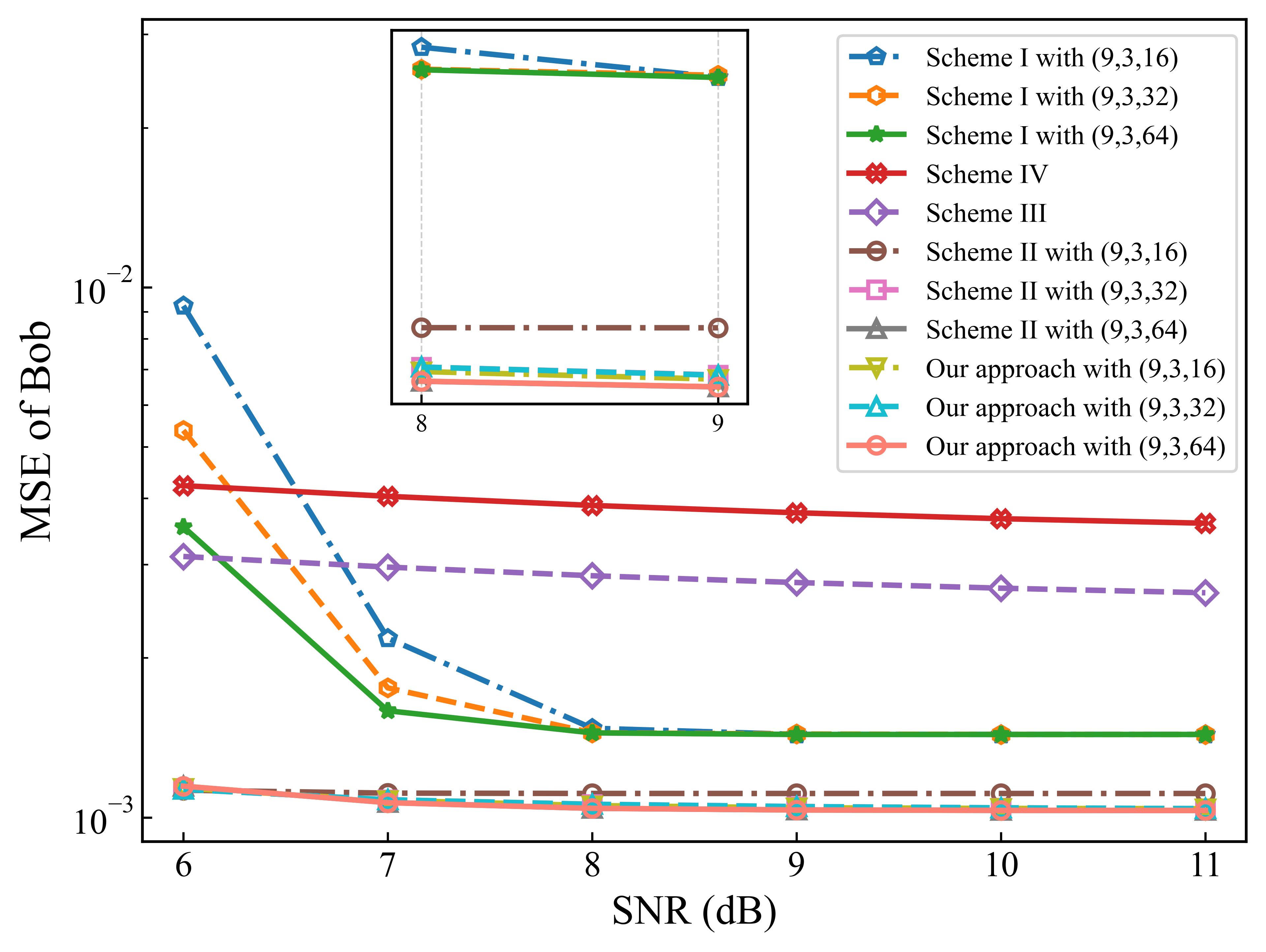}
			\label{fig:mse_bob}}
		\hfil
		\subfloat[]{%
			\includegraphics[width=0.45\linewidth]{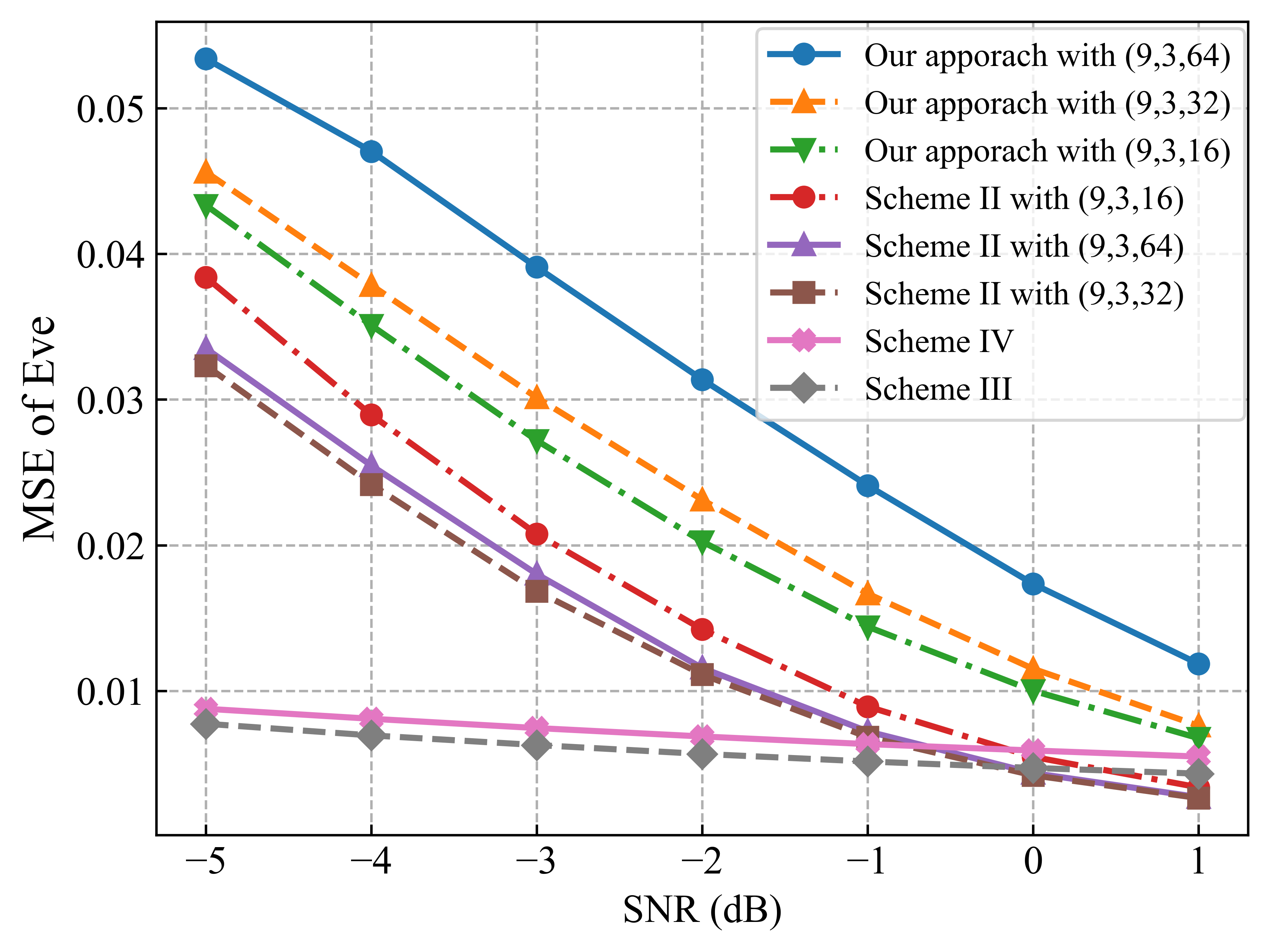}
			\label{fig:mse_eve}}
		\caption{{\color{blue}Recovery performance (measured by MSE) versus SNR for (a) Bob and (b) Eve, evaluated under different codeword lengths $n$ and their corresponding hyperparameter configurations $(\alpha_1, \alpha_2, \alpha_3)$: $n=16$ with $(1, 1.7, 1)$, $n=32$ with $(1, 7.5, 2)$, and $n=64$ with $(1, 20, 4)$.}}
		\label{fig:mse_bob_eve}
	\end{figure*}
	
	\begin{figure*}[!htb]
		\centering
		\subfloat{\includegraphics[width=0.45\linewidth]{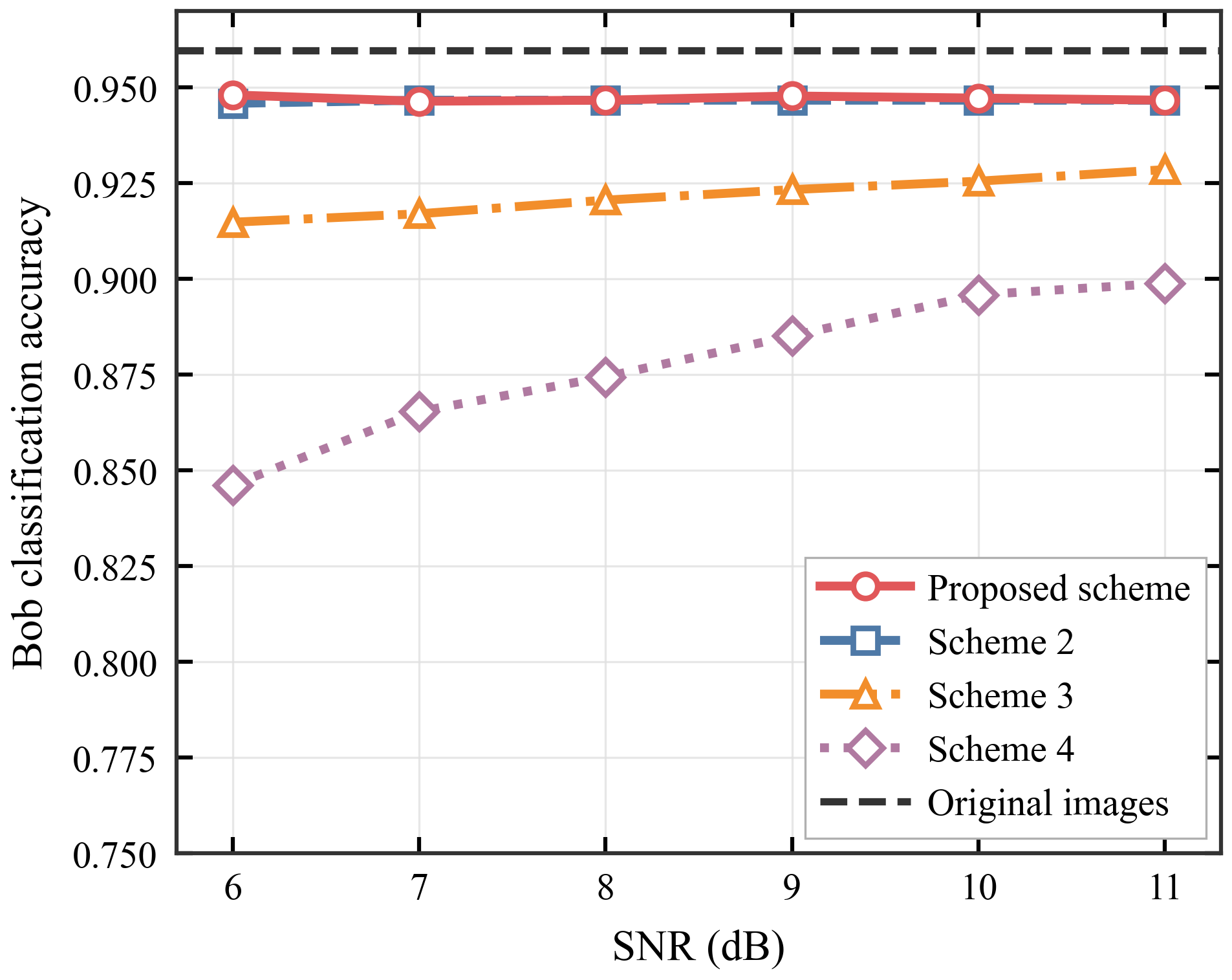}}
		\hfill
		\subfloat{\includegraphics[width=0.45\linewidth]{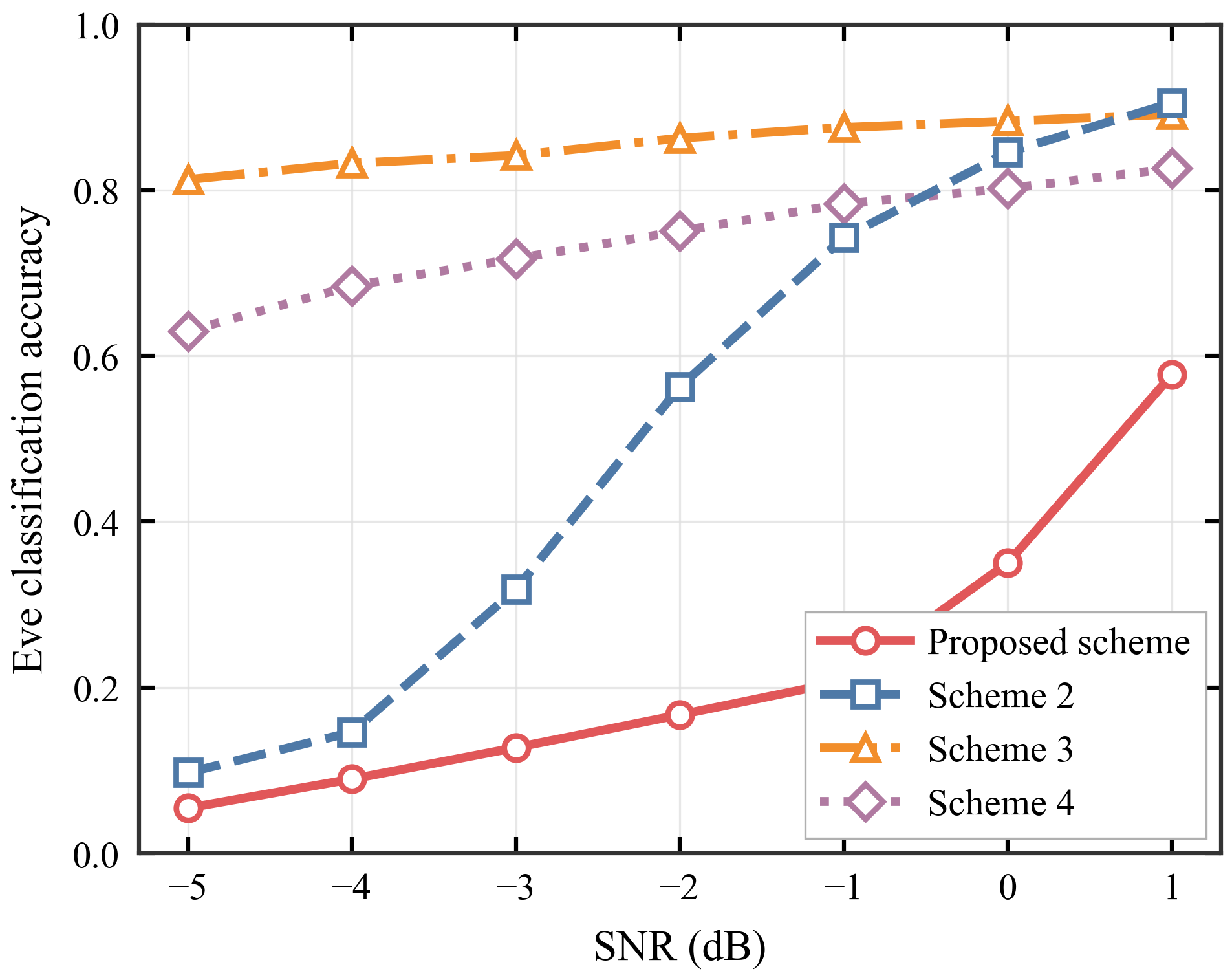}}
		\caption{{\color{blue}Downstream classification accuracy versus SNR at Bob and Eve under $(g, r, n) = (9, 3, 64)$.}}
		\label{fig:classifier_accuracy_optimal_decoder}
	\end{figure*}
	\subsubsection{Validation of Information Leakage Estimator and Upper Bound}
	
		{\color{blue}Fig.~\ref{fig:leakage_bound_validation} compares the benchmark information leakage against the CLUB-based estimator and the proposed coset-structure-based upper bound. In this validation experiment, the information block $\mathbf{Z}$ is assumed to follow a uniform distribution over $\{0,1\}^k$, offering a well-defined reference for leakage magnitudes; under this setup, the maximum leakage equals $H(\mathbf{Z})/n=k/n$ bits per channel use. To maintain tractability for numerically evaluating the capacity term within the proposed upper bound, we adopt the setting $(g,r,n)=(3,1,5)$. For this validation experiment, the admissible channel-input support is taken to contain $2^g=8$ distinct codewords within the length-$n$ binary input space of cardinality $2^n=32$, leading to $\binom{32}{8}$ feasible input-set configurations. The benchmark curve is independently constructed via input-block enumeration combined with Monte Carlo averaging over channel observations, and all leakage metrics are normalized by $n\ln 2$. As depicted, the CLUB-based estimator and the proposed upper bound both capture the monotonically rising trend of information leakage as Eve's SNR increases.  Notably, the CLUB-based estimator exhibits closer 
		agreement with the benchmark in the low-SNR region, whereas the proposed 
		upper bound demonstrates superior tightness in the high-SNR region. At higher Eve SNRs, CLUB-based estimates surpass the plotted upper bound \(H(\mathbf{Z})/n=0.4\) bits per channel use and are omitted to maintain readability.}
	
	{\color{blue}The aforementioned performance discrepancies primarily stem from the distinct evolution of the 
		respective relaxation factors. For the CLUB-based estimator, the deviation  is primarily determined by the Jensen gap and the variational approximation error. When the conditional density $p(\mathbf{z}_{\mathrm{E}} | \mathbf{z})$ varies slightly with $\mathbf{z}$, Jensen's inequality approaches equality. In the low-SNR regime, the larger noise variance weakens the dependence of the channel output $\mathbf{z}_{\mathrm{E}}$ on the information block $\mathbf{z}$, making the conditional distributions associated with different information blocks more similar. This reduces the Jensen gap and improves the agreement between the CLUB-based estimator and the benchmark leakage.}
	{\color{blue}In contrast, the tightness of the proposed coset-structure-based bound is mainly determined by the gap between the capacity term $\mathcal{C}_{\mathrm E}$ and the actual mutual information achieved. At high SNRs, mutual information under the capacity-achieving distribution and the codebook-induced distribution approaches the entropy of the corresponding input distribution. Consequently, the gap between these two mutual information quantities shrinks when the corresponding input entropies are close, yielding a tighter bound.}

	\subsubsection{Performance Visualization on the Kodak Dataset} 
	The visual results in Fig.~\ref{fig:kodak_visual_comparison} compare the schemes at two SNR levels at Bob, namely  10~dB   and 5~dB, corresponding to relatively favorable and degraded channel conditions, respectively. At 10~dB, the proposed scheme achieves an MSE of 0.0002  and an FCS of 0.9977, outperforming Scheme~III (0.0007/0.9868) and Scheme~IV (0.0022/0.9593) in both distortion and similarity. {\color{blue}When Bob's SNR decreases  to 5~dB, the proposed scheme   retains  an MSE of 0.0003   and an FCS of 0.9950. It outperforms Scheme~I (0.0115/0.9248), Scheme~III (0.0010/0.9781), and Scheme~IV (0.0028/0.9318), while achieving performance  comparable to Scheme~II. Although Schemes~I and  II   both employ  PWC for channel coding, the   performance advantage of Scheme~II suggests that learned semantic features are less sensitive  to channel distortion   than conventionally compressed source data under noisy channel conditions.}
	
	{\color{blue} For the security evaluation, we consider Eve's SNR of $-3$~dB. The proposed scheme   yields  an MSE of 0.0170   and an FCS of 0.1249 at Eve, with little recognizable content retained in the recovered image. By comparison, Scheme~II achieves an  MSE of 0.0088   and an FCS of 0.6277, while Schemes~III and IV exhibit substantially higher FCS values of 0.9594  and   0.8894, respectively. In particular, the high FCS obtained by Scheme~III indicates that its recovered image remains close to the source image in the DINOv2 feature space. These results show that the proposed scheme more effectively limits the information recoverable by Eve.}
	{\color{blue}Scheme~I is not included in this comparison because its deterministic JPEG decoder cannot be optimized for Eve's channel conditions.}

	\subsubsection{Experimental Performance Analysis on the ImageNet and Moving MNIST Datasets}

	Fig.~\ref{fig:mse_bob_eve}(a) compares Bob's MSE over SNRs  ranging from 6 to 11~dB. {\color{blue}For Scheme~I, the  codeword length of 64   yields the lowest MSE, whereas the length of  16   gives the highest, showing the reliability benefit of a longer codeword. Nevertheless, Scheme~I exhibits higher MSEs at 6--7~dB, indicating greater sensitivity to channel noise.}   Schemes~III and IV   provide  more stable performance   over the considered  SNR range,   although their MSEs decrease only marginally at higher SNRs. {\color{blue}The proposed scheme achieves lower MSE than Schemes~I, III, and IV   throughout the evaluated range, reaching approximately $1\times10^{-3}$  at 9~dB.}

	{\color{blue}On Eve's side, Fig.~\ref{fig:mse_bob_eve}(b) reports Eve's MSE to characterize the security performance of the evaluated  schemes. For Schemes~III and IV,   Eve's recovery error remains below $1\times10^{-2}$ even at $-5$~dB, indicating that these schemes provide limited protection against eavesdropping. By  comparison, Scheme~II yields higher recovery errors, with the gap becoming more pronounced as Eve's SNR decreases.} However, the security performance of PWC does not   improve monotonically with  the codeword length. {\color{blue}This non-monotonic behavior arises from the channel-polarization mechanism, which classifies the polarized subchannels according to  their reliability. As the codeword length   changes, the resulting reliability ordering and channel allocation are also altered, affecting the balance  between highly reliable   subchannels used for information transmission and weaker subchannels used to introduce uncertainty.}
	{\color{blue}The proposed scheme yields consistently higher Eve-side recovery errors than Scheme~II. In particular, the configuration  with a codeword length of 64   produces the largest recovery error at  Eve among all   evaluated settings.}

	{\color{blue}Fig.~\ref{fig:classifier_accuracy_optimal_decoder} further evaluates downstream task performance through image classification accuracy. The recovered samples are fed into the pretrained classifier, and the resulting accuracy reflects the extent to which discriminative category-level semantics are preserved. The proposed scheme maintains a classification accuracy of approximately 95\% at Bob, comparable to that of Scheme~II. At Eve, its substantially lower accuracy suggests weaker category-level semantic leakage, consistent with the MSE-based evaluation. For Schemes~III and IV, Eve's accuracy remains above 60\% even at $-5$~dB, indicating stronger leakage under this metric.}

	{\color{blue}These results further highlight the distinct security behaviors of continuous- and discrete-valued representations. For schemes using continuous representations, Eve's semantic recovery performance varies smoothly with the SNR, consistent with the graceful-degradation behavior of continuous-valued transmission~\cite{visual_protection_jscc}. Although this property supports robust recovery performance at Bob, it may also allow Eve to retain non-negligible information when the eavesdropping channel is only moderately degraded. By contrast, when Bob maintains a channel advantage, the proposed discrete representation preserves reliable transmission at Bob while causing Eve's performance to decline more sharply as the eavesdropping channel deteriorates. This discrepancy stems from  the   distinct noise-induced distortion patterns of continuous latent features and discrete symbol sequences. Channel noise perturbs continuous representations incrementally, while discrete symbol errors trigger sharp discontinuities in the transmitted information embedding, which injects far higher ambiguity into the semantic reconstruction recovered by the eavesdropper.} 
	
	\begin{table*}[!t]
		\centering
		\renewcommand{\arraystretch}{1.2}
		\resizebox{0.9\textwidth}{!}{\begin{tabular}{|l|c|c|c|c|c|c|c|c|c|c|}
				\hline
				\multirow{2}{*}{Methods} & \multicolumn{5}{c|}{\textbf{SNR for Eve}} & \multicolumn{5}{c|}{\textbf{SNR for Bob}} \\ \cline{2-11} 
				& -5~dB & -4~dB & -3~dB & -2~dB & -1~dB & 6~dB & 7~dB & 8~dB & 9~dB & 10~dB \\ \hline
				Scheme~I & N/A & N/A & N/A & N/A & N/A & 0.0159 & 0.0015 & 0.0002 & 0.0002 & 0.0002 \\ \hline
				Scheme~II & 0.0239 & 0.0147 & 0.0057 & 0.0027 & 0.0008 & 0.0001 & 0.0001 & 0.0001 & 0.0001 & 0.0001 \\ \hline
				Scheme~III & 0.0020 & 0.0015 & 0.0012 & 0.0009 & 0.0007 & 0.0002 & 0.0002 & 0.0001 & 0.0001 & 0.0001 \\ \hline
				Scheme~IV & 0.0030 & 0.0022 & 0.0012 & 0.0009 & 0.0007 & 0.0003 & 0.0003 & 0.0003 & 0.0003 & 0.0003 \\ \hline
				Proposed scheme & 0.0310 & 0.0208 & 0.0122 & 0.0055 & 0.0022 & 0.0001 & 0.0001 & 0.0001 & 0.0001 & 0.0001 \\ \hline
		\end{tabular}}
		\caption{{\color{blue}Video transmission performance (measured by MSE) versus SNR  on the Moving MNIST dataset with   $(11, 4, 32)$.}}
		\label{tab:moving_mnist_mse}
	\end{table*}
	
	\begin{figure}[!t]
		\centering
		\includegraphics[width=0.9\linewidth]{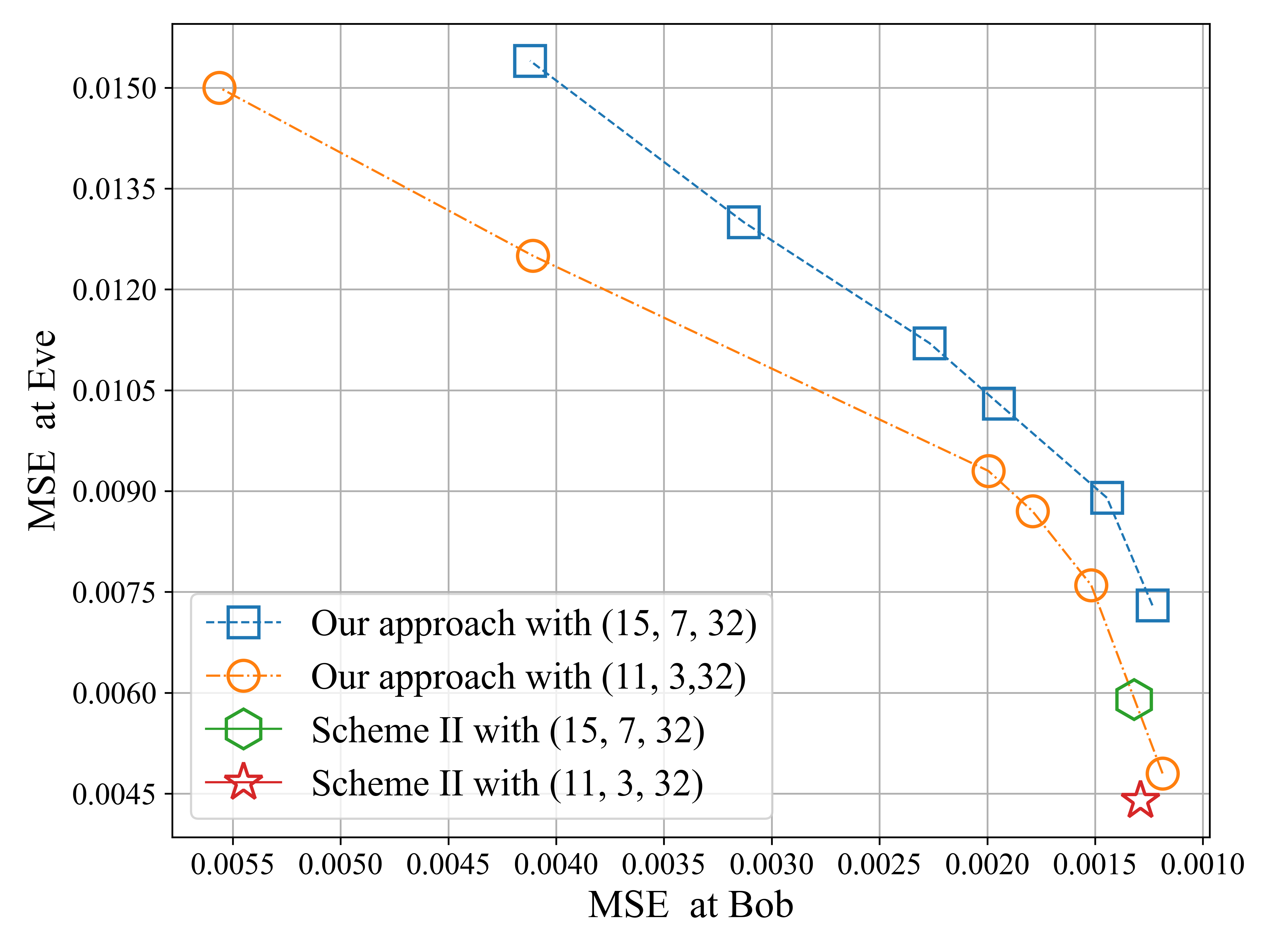}
		\caption{{\color{blue}Reliability-security trade-off (measured by MSE).}}
		\label{fig:reliability_security_tradeoff}
	\end{figure}
	\begin{figure*}[!t]
		\centering
		\subfloat[]{%
			\includegraphics[width=0.45\linewidth]{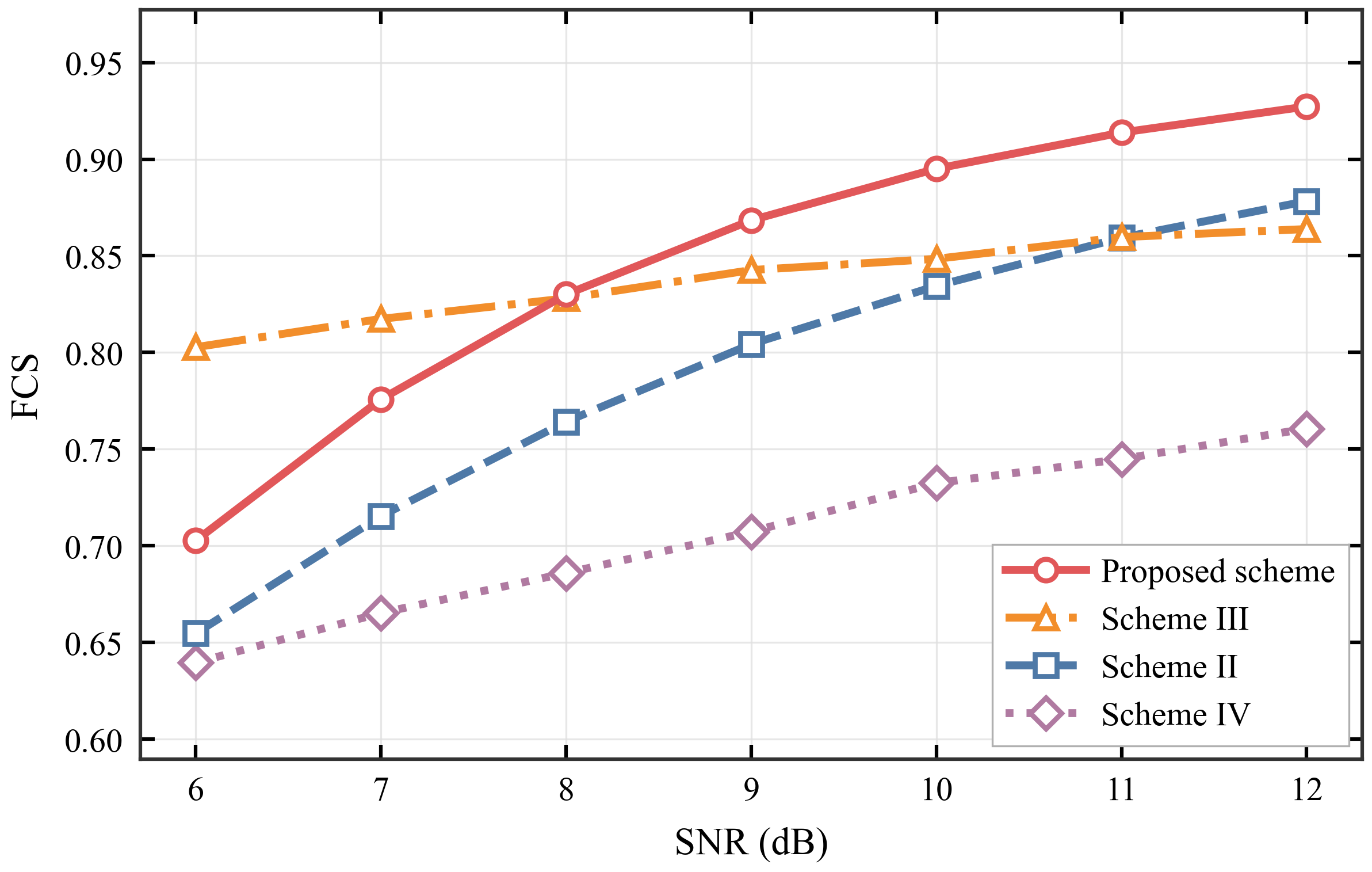}}
		\hfil
		\subfloat[]{%
			\includegraphics[width=0.45\linewidth]{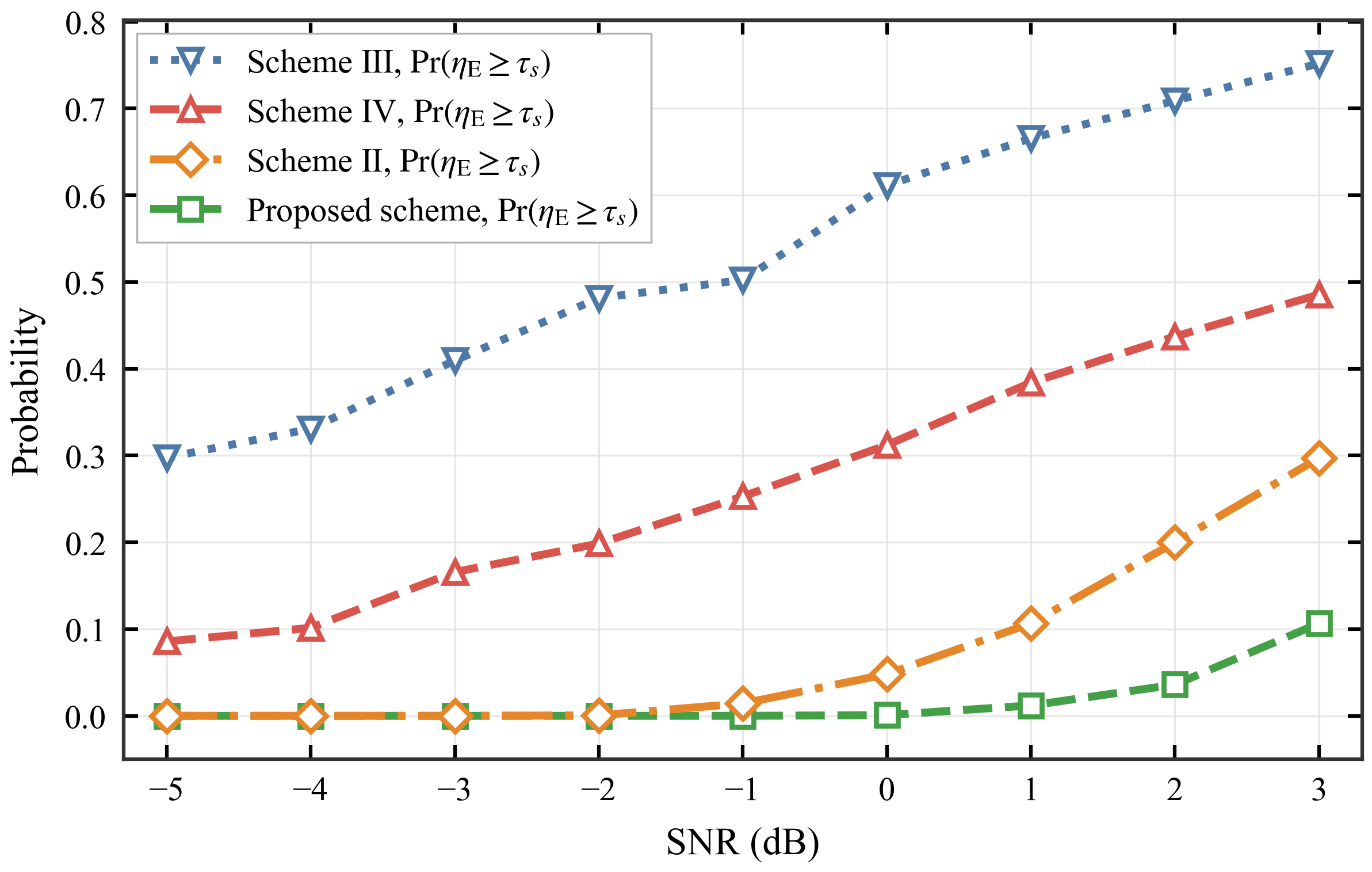}}
		\caption{{\color{blue}Performance under the block-fading channel: (a) average FCS at Bob and (b) semantic leakage probability at Eve.}}
		\label{fig:similarity_outage_probability}
	\end{figure*}

	\begin{figure*}[!htb]
		\centering
		\includegraphics[width=1.0\linewidth]{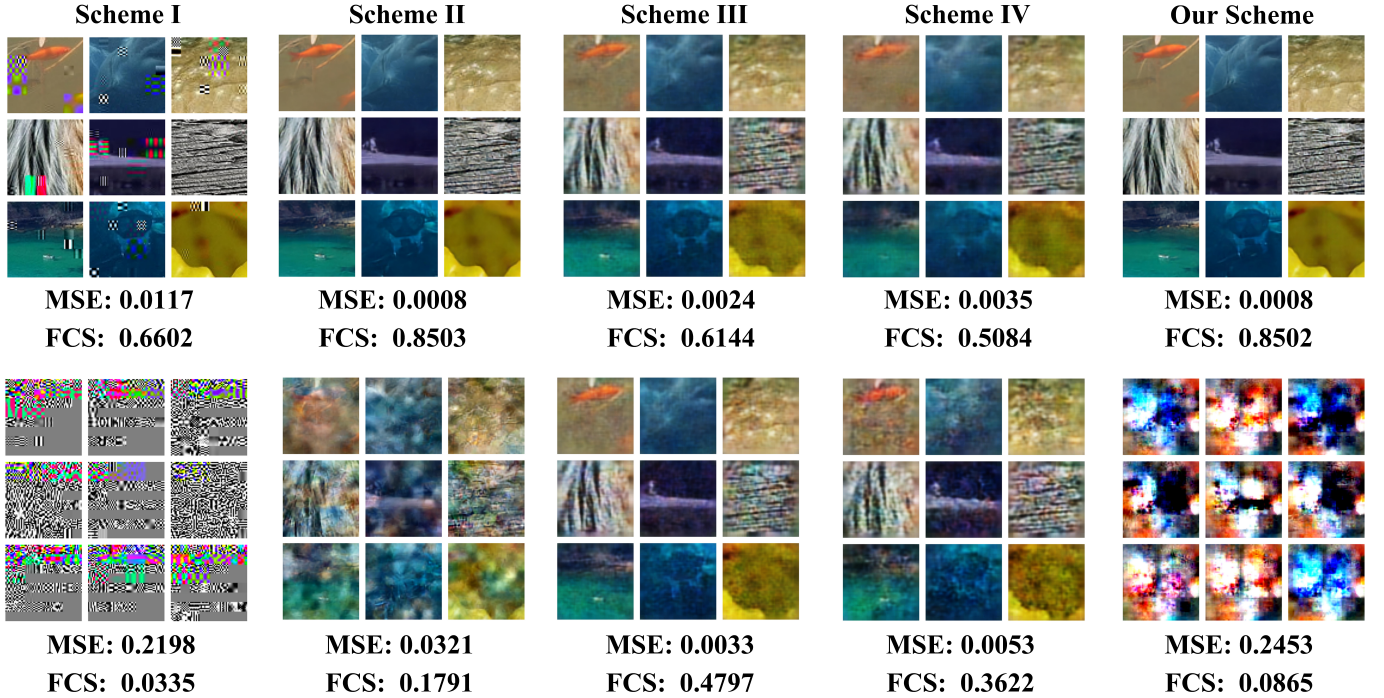}
		\caption{{\color{blue}Visual comparison   between  the proposed method and four baselines on  the  ImageNet   dataset,   under the encoding parameters $(g, r, n) = (36, 22, 64)$.   The first and second rows display the recovered images  at Bob (5~dB) and Eve (0~dB), respectively. The hyperparameters are configured as $\beta_1 = 6$, $\beta_2 = 0.087$, and $\beta_3 = 0.015$.}}
		\label{fig:visual_identical_decoder}
	\end{figure*}
	
	\begin{figure}[!htb]
		\centering
		\includegraphics[width=0.9\linewidth]{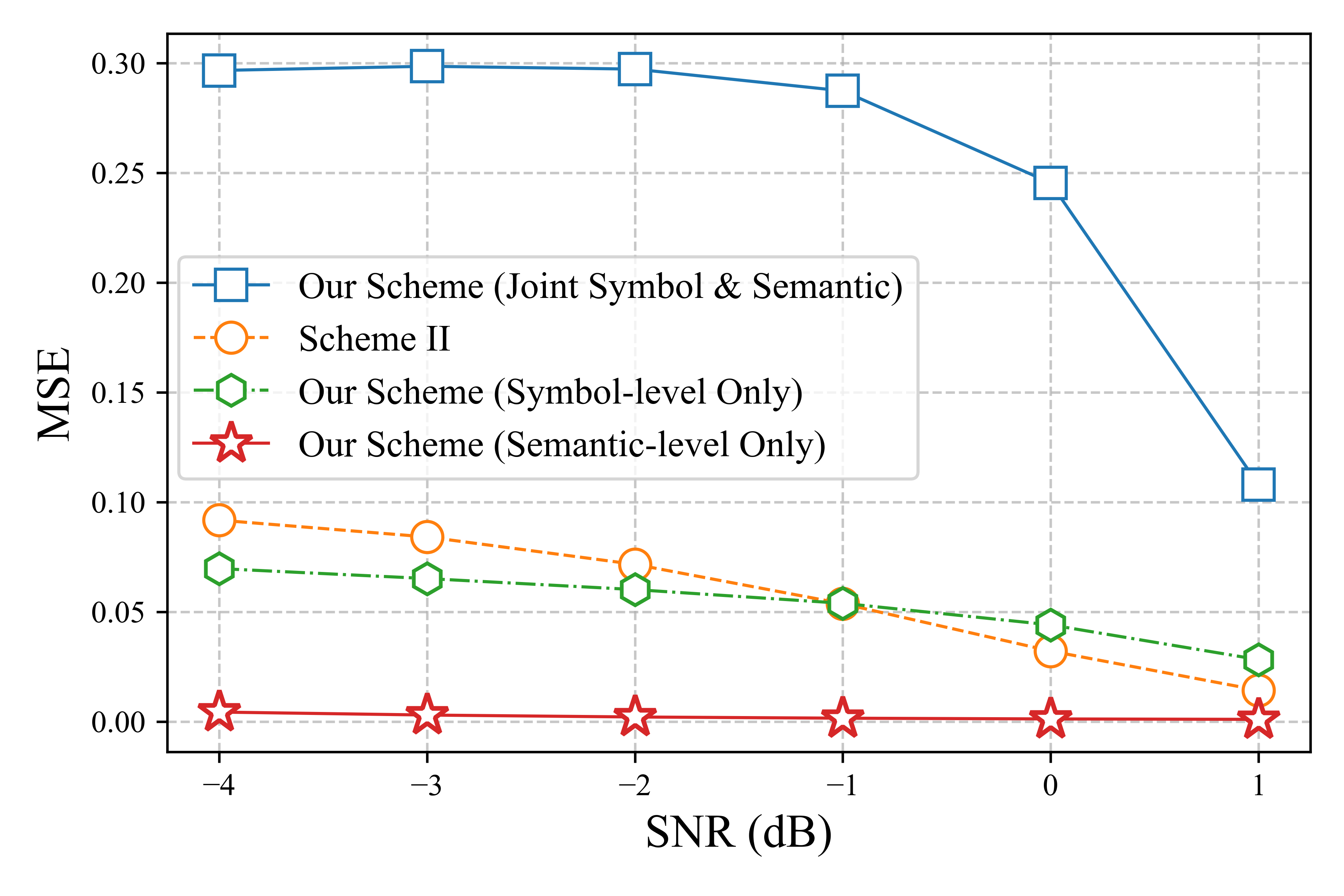}
		\caption{{\color{blue}Eve-side recovery performance (measured by MSE) versus SNR under $(36, 22, 64)$.}}
		\label{fig:symbol_semantic_optimization}
	\end{figure}
	
	{\color{blue}We further verify the effectiveness of the proposed scheme in a video transmission scenario   using  the Moving MNIST dataset. As shown in Table~\ref{tab:moving_mnist_mse}, similar trends are observed. Scheme~I suffers from substantial performance degradation at low SNRs, while Scheme~II and the proposed scheme maintain low  errors and stable decoding performance at Bob's side. On Eve's side, the proposed method consistently yields the highest reconstruction error, again   indicating stronger  confidentiality. These results confirm that the   security-reliability trade-off  achieved in the image transmission task also holds for video transmission settings.}
	
	\subsubsection{Analyzing Trade-offs Between Reliability and Security}
	This experiment   investigates the flexibility of our  proposed scheme in   adjusting the reliability-security  trade-off, with the results  illustrated in Fig.~\ref{fig:reliability_security_tradeoff}. {\color{blue}The horizontally reversed  x-axis   represents Bob's MSE at  4~dB, whereas  the y-axis   represents Eve's decoding distortion at  0~dB. Accordingly, points located closer to the upper-right region correspond to lower distortion at Bob and higher distortion at Eve.}
	
	The hexagram  and star markers   represent the fixed operating points of Scheme~II under the $(15, 7, 32)$ and $(11, 3, 32)$ configurations,   respectively. By contrast, the  square and circle markers show that the proposed scheme   produces multiple operating  points under the   corresponding settings by varying the weights in the  training objective. {\color{blue}For example,   under the $(15, 7, 32)$ configuration, the operating point varies from $(0.0012, 0.0073)$ to $(0.0041, 0.0154)$, whereas under the $(11, 3, 32)$ configuration, it varies from $(0.0012, 0.0048)$ to $(0.0056, 0.0150)$. These results indicate that the proposed objective enables the reliability-security operating point to be adjusted according to different reliability-security priorities~\cite{a17}.}

	\subsubsection{Evaluation in Block-Fading Channels}
	
		{\color{blue}In wireless environments,  the   relative ordering of the instantaneous SNRs at Bob and Eve may vary across transmission blocks. To evaluate the proposed scheme under channel conditions beyond the idealized degraded AWGN assumption, we introduce a block-fading model. In this evaluation setup, Alice's encoder and Bob's decoder utilize parameters trained under AWGN channels and undergo no further adaptation over fading channels, whereas Eve's decoder is re-optimized using channel observations collected under fading realizations, consistent with the optimal-decoder Eve evaluation setting.  In this setting, we assume the legitimate link maintains an average statistical advantage over the wiretap channel, though instantaneous non-degraded channel realizations may still arise. This experiment investigates the performance under such channel variations, while a comprehensive theoretical analysis of fading wiretap channels lies beyond the scope of this work.

		Specifically, for the $\ell$-th block, the received signals at Bob and Eve are modeled, respectively, as
		\begin{equation}
			\mathbf{Z}_{\mathrm{B}}^{(\ell)}
			=
			h_{\mathrm{B}}^{(\ell)}
			\mathcal{M}\!\left(\mathbf{S}^{(\ell)}\right)
			+
			\boldsymbol{\epsilon}_{\mathrm{B}}^{(\ell)},
		\end{equation}
		and
		\begin{equation}
			\mathbf{Z}_{\mathrm{E}}^{(\ell)}
			=
			h_{\mathrm{E}}^{(\ell)}
			\mathcal{M}\!\left(\mathbf{S}^{(\ell)}\right)
			+
			\boldsymbol{\epsilon}_{\mathrm{E}}^{(\ell)},
		\end{equation}
		where $\mathcal{M}(\cdot)$ denotes the BPSK modulation function. {\color{blue}The channel coefficients of the two links are independent, with $h_{\mathrm{B}}^{(\ell)}\sim\mathcal{CN}(0,1),\qquad h_{\mathrm{E}}^{(\ell)}\sim\mathcal{CN}(0,1)$ across blocks; each coefficient remains constant within one block. Bob and Eve are assumed to know their own instantaneous channel coefficients and perform coherent equalization before decoding. 
		
		To examine security under the block-fading setting, we primarily evaluate the semantic leakage probability
		\begin{align}
			P_{\mathrm{leak}}=\Pr(\eta_{\mathrm{E}}\geq\tau_s),
		\end{align}
		where $\eta_{\mathrm{E}}$ denotes the FCS value at Eve and $\tau_s$ denotes the semantic leakage threshold. Specifically, $\tau_s$ is determined as the 99th percentile of cross-class maximum FCS values over the validation set, yielding $\tau_s=0.6765$ in our evaluation. A semantic leakage event is declared when Eve's FCS exceeds this threshold, where the recovered content preserves category-consistent source features and thus exceeds the typical cross-class reference level.}
		
		Fig.~\ref{fig:similarity_outage_probability}(a) depicts the average FCS at Bob across different average SNRs. Similar to the observations under AWGN channels, Scheme~III maintains comparatively high similarity in the 6--7~dB range. Benefiting from continuous representations and the absence of semantic-security constraints, it experiences relatively gentle performance degradation under channel fluctuations. In contrast, the proposed scheme improves steadily as Bob's average SNR increases and achieves the highest similarity from 8~dB onward, indicating reliable semantic recovery over block-fading channels at moderately high SNRs despite mild degradation at 6--7~dB. Fig.~\ref{fig:similarity_outage_probability}(b) plots \(P_{\mathrm{leak}}\) against Eve's average SNR over block-fading channels. As Eve's average SNR increases, the leakage probability of all schemes tends to increase. For Scheme~IV, \(P_{\mathrm{leak}}\) reaches approximately \(0.44\) at an average SNR of 2~dB and exceeds \(0.49\) at 3~dB, whereas the proposed scheme remains around \(10.7\%\) at 3~dB. These clear gaps in semantic leakage probability further illustrate that the proposed scheme can suppress semantic leakage under fading-induced channel variations.}

	\subsection{Performance Evaluation with the   Identical  Decoder for Eve and Bob}
	This subsection   considers  the scenario where   the  eavesdropper, Eve,   employs the same decoding instance as the legitimate receiver, Bob.
	
	Fig.~\ref{fig:visual_identical_decoder}  provides a visual comparison between the proposed scheme and four baseline   schemes  on the ImageNet dataset, with the top and bottom rows showing the recovered  images at Bob   (5~dB) and Eve (0~dB), respectively. {\color{blue}At Bob,  the proposed scheme and Scheme~II achieve comparable performance, with MSE/FCS values  of 0.0008/0.8502 and 0.0008/0.8503, respectively, indicating similar recovery quality. Scheme~I exhibits more pronounced visual artifacts, with an MSE of 0.0117 and an FCS of 0.6602, reflecting its greater sensitivity to  channel impairments. Schemes~III and IV also   show degraded Bob-side performance, yielding MSE/FCS values of 0.0024/0.6144 and 0.0035/0.5084, respectively.}
	
	At Eve,  the   differences among the evaluated schemes become more pronounced. {\color{blue}Scheme~II yields an MSE of 0.0321 and an FCS of 0.1791, indicating that little source-related visual information remains discernible. Schemes~III (0.0033/0.4797) and IV (0.0053/0.3622) exhibit greater semantic leakage, as their decoded content retains higher similarity to the source images. Scheme~I yields a low FCS of 0.0335 at Eve, but also suffers substantial degradation at Bob. This behavior reflects the cliff effect of conventional separation-based transmission, whereby recovery performance can deteriorate abruptly under moderately degraded channel conditions. Although this characteristic can restrict the information recoverable by an eavesdropper with an inferior channel, it also poses a reliability challenge when the legitimate receiver experiences unfavorable channel conditions.}  {\color{blue}In contrast, the proposed scheme   maintains reliable recovery at Bob while markedly reducing Eve's recovery quality, yielding an MSE of 0.2453 and an FCS of 0.0865 at Eve. The recovered images contain little information, and these  results indicate a favorable balance between reliable communication for Bob and reduced semantic leakage at Eve.}
	
	{\color{blue}To examine the respective roles of the two GMI-based security objectives, we conduct   an ablation study  under three optimization configurations: 
		(i) symbol-level optimization, which minimizes   $\mathrm{GMI}(\mathbf{Z};\mathbf{Z}_{\mathrm{E}})$ to reduce Eve's symbol-level recoverability under the prescribed decoder; 
		(ii) semantic-level optimization, which minimizes   $\mathrm{GMI}(\mathbf{Y};\mathbf{V}_{\mathrm{E}})$ to limit Eve's decoder-specific semantic recoverability; and 
		(iii) joint optimization, which   incorporates both objectives.} 
	{\color{blue}Throughout this simulation, we evaluate different optimization setups under the constraint that Bob's recovery performance remains comparable across all cases.}

	As shown in Fig.~\ref{fig:symbol_semantic_optimization},  at the SNR of   $-1$~dB,  the symbol-level and semantic-level variants   yield  MSEs of 0.052 and 0.0039, respectively. {\color{blue}In contrast,   jointly incorporating the two objectives increases the Eve-side MSE  to 0.286, indicating a   pronounced degradation in the recovered content. Notably,  when Bob and Eve   are subject to comparable  channel conditions,   the semantic feature vectors they recover tend to exhibit strong consistency. In this case, semantic-only optimization fails to create a clear gap between their performance. By contrast, the introduction of  symbol-level   suppression directly impairs the quality of Eve's recovered semantic features. This allows the semantic optimization module to further limit the amount of  semantic information Eve can extract. As such, the two optimization targets deliver complementary secrecy protection.}

	\section{Conclusion}
	This paper investigated a wiretap coding scheme within  the semantic communication framework, with the objective of maintaining reliable transmission to the legitimate receiver while limiting the information recoverable by the eavesdropper. {\color{blue}Two eavesdropper models with strong and weak assumptions on Eve's decoding capability were further considered. These models characterize representative ways in which Eve may acquire effective decoding functionality, and a customized secure coding strategy was developed for each setting. By integrating structured coset coding into semantic communication and adopting information-theoretic reliability and leakage bounds as tractable surrogate objectives, the proposed framework realizes information-theoretically guided end-to-end optimization. This distinguishes our method from conventional adversarial training and privacy-constrained optimization.} Variational  approximation, contrastive log-ratio approximation, and Monte Carlo estimation   were employed to obtain tractable estimates of these objectives. {\color{blue}Simulation results   showed that the proposed schemes outperformed the considered separation-based baselines and end-to-end methods, achieving higher reliability at the legitimate receiver while more effectively limiting the information recoverable by the eavesdropper. The proposed framework also supports flexible adjustment of the trade-off between  transmission reliability and confidentiality   according to different system requirements. Future work will explore theoretical and algorithmic extensions for general non-degraded, time-varying wireless environments, expanding the applicability of the proposed methodology beyond the idealized static degraded wiretap channel setup.}
	
	\bibliographystyle{IEEEtran}
	\bibliography{ref.bib}

@inproceedings{2024wiretap,
		author    = {Xiangnan Zhou and Chao Wang and Haibin Zhang and Yao Sun and Chonghua Wang and Derrick Wing Kwan Ng},
		title     = {Practical Wiretap Code Design via Integration of Information Theory and Deep Learning},
		booktitle = {Proc. IEEE Global Commun. Conf. (GLOBECOM) Workshops},
		year      = {2024},
		pages     = {1--6},
		doi       = {10.1109/GCWkshp64532.2024.11101598}
	}

@ARTICLE{a1,
	author={Lou, Wenjing},
	journal={IEEE Wireless Commun.}, 
	title={Security, privacy, and accountability in wireless access networks}, 
	year={2009},
	volume={16},
	number={4},
	pages={80--87},
	doi={10.1109/MWC.2009.5281259}}

@article{a2,
	title={A Review and Comparative Analysis of Various Encryption Algorithms},
	author={Rajdeep Bhanot and Rahul Hans},
	journal={Int. J. Secur. Appl.},
	year={2015},
	volume={9},
	number={4},
	pages={289--306},
	doi={10.14257/ijsia.2015.9.4.27},
}

@ARTICLE{a6,
	author={Wu, Yongpeng and Khisti, Ashish and Xiao, Chengshan and Caire, Giuseppe and Wong, Kai-Kit and Gao, Xiqi},
	journal={IEEE J. Sel. Areas Commun.}, 
	title={A Survey of Physical Layer Security Techniques for {5G} Wireless Networks and Challenges Ahead}, 
	year={2018},
	volume={36},
	number={4},
	pages={679--695},
	doi={10.1109/JSAC.2018.2825560},
}

@ARTICLE{endogenous1,
	author={Jin, Liang and Hu, Xiaoyan and Lou, Yangming and Zhong, Zhou and Sun, Xiaoli and Wang, Huiming and Wu, Jiangxing},
	journal={China Commun.}, 
	title={Introduction to wireless endogenous security and safety: Problems, attributes, structures and functions}, 
	year={2021},
	volume={18},
	number={9},
	pages={88--99},
	doi={10.23919/JCC.2021.09.008}}

@article{endogenous2,
	title={Discussion on a New Paradigm of Endogenous Security Towards {6G} Networks},
	author={Ji, Xinsheng and Wu, Jiangxing and Jin, Liang and Huang, Kaizhi and Chen, Yajun and Sun, Xiaoli and You, Wei and Huo, Shumin and Yang, J.},
	journal={Front. Inf. Technol. Electron. Eng.},
	year={2022},
	volume={23},
	number={10},
	pages={1421--1450},
	doi={10.1631/FITEE.2200385}}

@article{a21,
	title={The {Gaussian} Wire-Tap Channel},
	author={Leung-Yan-Cheong, Sik K. and Hellman, Martin E.},
	journal={IEEE Trans. Inf. Theory},
	volume={24},
	number={4},
	pages={451--456},
	year={1978},
	doi={10.1109/TIT.1978.1055917}}

@article{wyner,
	title={The wire-tap channel},
	author={Wyner, Aaron D},
	journal={Bell Syst. Tech. J.},
	volume={54},
	number={8},
	pages={1355--1387},
	year={1975},
	doi={10.1002/j.1538-7305.1975.tb02040.x}
	}

@ARTICLE{a16,
	author={Mahdavifar, Hessam and Vardy, Alexander},
	journal={IEEE Trans. Inf. Theory}, 
	title={Achieving the Secrecy Capacity of Wiretap Channels Using Polar Codes}, 
	year={2011},
	volume={57},
	number={10},
	pages={6428--6443},
	doi={10.1109/TIT.2011.2162275}}

@ARTICLE{a13,
	author={Thangaraj, Andrew and Dihidar, Souvik and Calderbank, A. R. and McLaughlin, Steven W. and Merolla, Jean-Marc},
	journal={IEEE Trans. Inf. Theory}, 
	title={Applications of {LDPC} Codes to the Wiretap Channel}, 
	year={2007},
	volume={53},
	number={8},
	pages={2933--2945},
	doi={10.1109/TIT.2007.901143}
}

@ARTICLE{a17,
	author={Besser, Karl-Ludwig and Lin, Pin-Hsun and Janda, Carsten R. and Jorswieck, Eduard A.},
	journal={IEEE Trans. Inf. Forens. Secur.}, 
	title={Wiretap Code Design by Neural Network Autoencoders}, 
	year={2020},
	volume={15},
	pages={3374--3386},
	doi={10.1109/TIFS.2019.2945619}}

@article{a18,
	title={Learning End-to-End Codes for the {BPSK}-Constrained {Gaussian} Wiretap Channel},
	journal={Phys. Commun.},
	volume={46},
	pages={101282},
	year={2021},
	issn={1874-4907},
	doi={10.1016/j.phycom.2021.101282},
	author={Alireza Nooraiepour and Sina Rezaei Aghdam}}

@inproceedings{a19,
	author = {Fritschek, Rick and Schaefer, Rafael F. and Wunder, Gerhard},
	title = {Deep Learning Based Wiretap Coding via Mutual Information Estimation},
	year = {2020},
	isbn = {9781450380072},
	publisher = {Association for Computing Machinery},
	address = {New York, NY, USA},
	doi = {10.1145/3395352.3402654},
	booktitle = {Proceedings of the 2nd ACM Workshop on Wireless Security and Machine Learning},
	pages = {74--79},
	numpages = {6},
	location = {Linz, Austria},
	series = {WiseML '20}}

@InProceedings{a22,
	title = {Mutual Information Neural Estimation},
	author = {Belghazi, Mohamed Ishmael and Baratin, Aristide and Rajeshwar, Sai and Ozair, Sherjil and Bengio, Yoshua and Courville, Aaron and Hjelm, Devon},
	booktitle = {Proceedings of the 35th International Conference on Machine Learning},
	pages = {531--540},
	year = {2018},
	editor = {Dy, Jennifer and Krause, Andreas},
	volume = {80},
	series = {Proceedings of Machine Learning Research},
	month = {10--15 Jul},
	publisher = {PMLR},
	url = {https://proceedings.mlr.press/v80/belghazi18a.html}}

@ARTICLE{a20,
	author={Rana, Vidhi and Chou, R{\'e}mi A.},
	journal={IEEE Trans. Commun.}, 
	title={Short Blocklength Wiretap Channel Codes via Deep Learning: Design and Performance Evaluation}, 
	year={2023},
	volume={71},
	number={3},
	pages={1462--1474},
	doi={10.1109/TCOMM.2023.3237259}}

@ARTICLE{rana2025helper,
		author={Rana, Vidhi and Chou, R{\'e}mi A. and Kim, Taejoon},
		journal={IEEE Trans. Commun.},
		title={Helper-Assisted Coding for {Gaussian} Wiretap Channels: Deep Learning Meets {PhySec}},
		year={2025},
		volume={73},
		number={12},
		pages={13199--13213},
		doi={10.1109/TCOMM.2025.3624165}
	}

@INPROCEEDINGS{seifert2025modular,
		author={Seifert, Daniel and G{\"u}nl{\"u}, Onur and Schaefer, Rafael F.},
		booktitle={Proc. IEEE Int. Symp. Personal, Indoor Mobile Radio Commun. (PIMRC)},
		title={Modular Neural Wiretap Codes for Fading Channels},
		year={2025},
		doi={10.1109/PIMRC62392.2025.11274860}
	}

@ARTICLE{Shannon,
	author={Shannon, C. E.},
	journal={The Bell System Technical Journal}, 
	title={A mathematical theory of communication}, 
	year={1948},
	volume={27},
	number={3},
	pages={379--423},
	doi={10.1002/j.1538-7305.1948.tb01338.x}}

@ARTICLE{Joint,
	author={G{\"u}nd{\"u}z, Deniz and Wigger, Mich{\`e}le A. and Tung, Tze-Yang and Zhang, Ping and Xiao, Yong},
	journal={Proc. IEEE}, 
	title={Joint Source-Channel Coding: Fundamentals and Recent Progress in Practical Designs}, 
	year={2025},
	volume={113},
	number={9},
	pages={888--919},
	doi={10.1109/JPROC.2024.3477331}}

@ARTICLE{JSCC,
	author={Bourtsoulatze, Eirina and Burth Kurka, David and G{\"u}nd{\"u}z, Deniz},
	journal={IEEE Trans. Cogn. Commun. Netw.}, 
	title={Deep Joint Source-Channel Coding for Wireless Image Transmission}, 
	year={2019},
	volume={5},
	number={3},
	pages={567--579},
	doi={10.1109/TCCN.2019.2919300}}

@book{weaver1949recent,
		author={Shannon, Claude E. and Weaver, Warren},
		title={The Mathematical Theory of Communication},
		publisher={University of Illinois Press},
		address={Urbana, IL, USA},
		year={1949}
	}

@article{xie2021deepSC,
		author={Xie, Huiqiang and Qin, Zhijin and Li, Geoffrey Ye and Juang, Biing-Hwang},
		title={Deep Learning Enabled Semantic Communication Systems},
		journal={IEEE Trans. Signal Process.},
		volume={69},
		pages={2663--2675},
		year={2021},
		doi={10.1109/TSP.2021.3071210}
	}

@article{gunduz2023beyond,
		author={G{\"u}nd{\"u}z, Deniz and Qin, Zhijin and Aguerri, Inaki Estella and Dhillon, Harpreet S. and Yang, Zhaohui and Yener, Aylin and Wong, Kai-Kit and Chae, Chan-Byoung},
		title={Beyond Transmitting Bits: Context, Semantics, and Task-Oriented Communications},
		journal={IEEE J. Sel. Areas Commun.},
		volume={41},
		number={1},
		pages={5--41},
		year={2023},
		doi={10.1109/JSAC.2022.3223408}
	}

@INPROCEEDINGS{Adversarial,
	author={Marchioro, Thomas and Laurenti, Nicola and G{\"u}nd{\"u}z, Deniz},
	booktitle={Proc. IEEE Int. Conf. Acoust., Speech Signal Process. (ICASSP)}, 
	title={Adversarial Networks for Secure Wireless Communications}, 
	year={2020},
	pages={8748--8752},
	doi={10.1109/ICASSP40776.2020.9053216}}

@INPROCEEDINGS{secure_deepjscc_multi_eve,
		author={Ameli Kalkhoran, Seyyed Amirhossein and Letafati, Mehdi and Erdemir, Ecenaz and Khalaj, Babak Hossein and Behroozi, Hamid and G{\"u}nd{\"u}z, Deniz},
		booktitle={Proc. IEEE Global Commun. Conf. (GLOBECOM)},
		title={Secure {Deep-JSCC} Against Multiple Eavesdroppers},
		year={2023},
		pages={3433--3438},
		doi={10.1109/GLOBECOM54140.2023.10436928}
	}

@ARTICLE{visual_protection_jscc,
		author={Xu, Jialong and Ai, Bo and Chen, Wei and Wang, Ning and Rodrigues, Miguel},
		journal={IEEE Trans. Cogn. Commun. Netw.},
		title={Deep Joint Source-Channel Coding for Image Transmission With Visual Protection},
		year={2023},
		volume={9},
		number={6},
		pages={1399--1411},
		doi={10.1109/TCCN.2023.3306851}
	}

@ARTICLE{privacy_aware_jscc,
		author={Letafati, Mehdi and Ameli Kalkhoran, Seyyed Amirhossein and Erdemir, Ecenaz and Khalaj, Babak Hossein and Behroozi, Hamid and G{\"u}nd{\"u}z, Deniz},
		journal={IEEE Trans. Mach. Learn. Commun. Netw.},
		title={Deep Joint Source Channel Coding for Privacy-Aware End-to-End Image Transmission},
		year={2025},
		volume={3},
		pages={568--584},
		doi={10.1109/TMLCN.2025.3564907}
	}

@article{du2022rethinking,
		author={Du, Hongyang and Wang, Jiacheng and Niyato, Dusit and Kang, Jiawen and Xiong, Zehui and Guizani, Mohsen and Kim, Dong In},
		title={Rethinking Wireless Communication Security in Semantic Internet of Things},
		journal={IEEE Wireless Commun.},
		year={2023},
		volume={30},
		number={3},
		pages={36--43},
		doi={10.1109/MWC.014.2200077}
	}

@article{privacy_multitask_semcom,
		author={Sagduyu, Yalin E. and Erpek, Tugba and Yener, Aylin and Ulukus, Sennur},
		title={Privacy-Preserving Semantic Communications via Multi-Task Learning and Adversarial Perturbations},
		journal={arXiv preprint arXiv:2512.24452},
		year={2025},
		doi={10.48550/arXiv.2512.24452}
	}

@ARTICLE{JSCA,
	author={Zhang, Maojun and Li, Yang and Zhang, Zezhong and Zhu, Guangxu and Zhong, Caijun},
	journal={IEEE Wireless Commun. Lett.}, 
	title={Wireless Image Transmission With Semantic and Security Awareness}, 
	year={2023},
	volume={12},
	number={8},
	pages={1389--1393},
	doi={10.1109/LWC.2023.3275383}}

@ARTICLE{DeepSSC,
	author={Li, Yongkang and Shi, Zheng and Hu, Han and Fu, Yaru and Wang, Hong and Lei, Hongjiang},
	journal={IEEE Commun. Lett.}, 
	title={Secure Semantic Communications: From Perspective of Physical Layer Security}, 
	year={2024},
	volume={28},
	number={10},
	pages={2243--2247},
	doi={10.1109/LCOMM.2024.3452715}}

@article{secure_semantic_wiretap,
		author={Kozlov, Denis and Mirmohseni, Mahtab and Tafazolli, Rahim},
		title={Secure Semantic Communication over Wiretap Channels: Rate-Distortion-Equivocation Tradeoff},
		journal={IEEE J. Sel. Areas Inf. Theory},
		year={2026},
		volume={7},
		pages={120--133},
		doi={10.1109/JSAIT.2026.3676713}
	}

@inproceedings{Club,
	author = {Cheng, Pengyu and Hao, Weituo and Dai, Shuyang and Liu, Jiachang and Gan, Zhe and Carin, Lawrence},
	title = {{CLUB}: A Contrastive Log-Ratio Upper Bound of Mutual Information},
	year = {2020},
	publisher = {JMLR.org},
	booktitle = {Proceedings of the 37th International Conference on Machine Learning},
	articleno = {166},
	numpages = {10},
	series = {ICML'20}}

@inproceedings{shuffle,
	author = {Sohn, Kihyuk},
	title = {Improved Deep Metric Learning With Multi-Class N-Pair Loss Objective},
	year = {2016},
	publisher = {Curran Associates, Inc.},
	booktitle = {Advances in Neural Information Processing Systems},
	volume = {29},
	pages = {1857--1865}
	}

@InProceedings{STE,
	author    = {Liu, Zechun and Cheng, Kwang-Ting and Huang, Dong and Xing, Eric P. and Shen, Zhiqiang},
	title     = {Nonuniform-to-Uniform Quantization: Towards Accurate Quantization via Generalized Straight-Through Estimation},
	booktitle = {Proc. IEEE/CVF Conf. Comput. Vis. Pattern Recognit. (CVPR)},
	month     = {June},
	year      = {2022},
	pages     = {4932--4942},
	doi       = {10.1109/CVPR52688.2022.00489}}

@inproceedings{costin2014firmware,
		author={Costin, Andrei and Zaddach, Jonas and Francillon, Aur{\'e}lien and Balzarotti, Davide},
		title={A Large-Scale Analysis of the Security of Embedded Firmwares},
		booktitle={Proc. 23rd USENIX Security Symposium (USENIX Security)},
		pages={95--110},
		year={2014}
	}

@ARTICLE{b20,
	author={Merhav, N. and Kaplan, G. and Lapidoth, A. and Shamai Shitz, S.},
	journal={IEEE Trans. Inf. Theory}, 
	title={On information rates for mismatched decoders}, 
	year={1994},
	volume={40},
	number={6},
	pages={1953--1967},
	doi={10.1109/18.340469}}

@book{b21,
	author={Fischer, Thomas R. M.},
	title="Some Remarks on the Role of Inaccuracy in Shannon's Theory of Information Transmission",
	bookTitle="Transactions of the Eighth Prague Conference: on Information Theory, Statistical Decision Functions, Random Processes held at Prague, from August 28 to September 1, 1978 Volume A",
	year="1978",
	address="Dordrecht",
	publisher="Springer",
	pages="211--226",
	isbn="978-94-009-9857-5",
	}

@article{oquab2023dinov2,
		author={Oquab, Maxime and Darcet, Timoth{\'e}e and Moutakanni, Th{\'e}o and Vo, Huy V. and Szafraniec, Marc and Khalidov, Vasil and Fernandez, Pierre and Haziza, Daniel and Massa, Francisco and El-Nouby, Alaaeldin and Assran, Mahmoud and Ballas, Nicolas and Galuba, Wojciech and Howes, Russell and Huang, Po-Yao and Li, Shang-Wen and Misra, Ishan and Rabbat, Michael and Sharma, Vasu and Synnaeve, Gabriel and Xu, Hu and J{\'e}gou, Herv{\'e} and Mairal, Julien and Labatut, Patrick and Joulin, Armand and Bojanowski, Piotr},
		title={{DINOv2}: Learning Robust Visual Features without Supervision},
		journal={Trans. Mach. Learn. Res.},
		year={2024}
	}

@INPROCEEDINGS{ImageNet,
	author={Deng, Jia and Dong, Wei and Socher, Richard and Li, Li-Jia and Kai Li and Li Fei-Fei},
	booktitle={Proc. IEEE Conf. Comput. Vis. Pattern Recognit. (CVPR)}, 
	title={{ImageNet}: A Large-Scale Hierarchical Image Database}, 
	year={2009},
	pages={248--255},
	doi={10.1109/CVPR.2009.5206848}}

@inproceedings{Moving,
	author = {Shi, Xingjian and Chen, Zhourong and Wang, Hao and Yeung, Dit-Yan and Wong, Wai-Kin and Woo, Wang-Chun},
	title = {Convolutional {LSTM} Network: A Machine Learning Approach for Precipitation Nowcasting},
	year = {2015},
	publisher = {Curran Associates, Inc.},
	booktitle = {Advances in Neural Information Processing Systems},
	volume = {28},
	pages = {802--810}
	}

@inproceedings{GDN,
	title={End-to-End Optimized Image Compression},
	author={Ball{\'e}, Johannes and Laparra, Valero and Simoncelli, Eero P.},
	booktitle={Proc. Int. Conf. Learn. Represent. (ICLR)},
	year={2017}
	}

@InProceedings{DVC,
	author = {Lu, Guo and Ouyang, Wanli and Xu, Dong and Zhang, Xiaoyun and Cai, Chunlei and Gao, Zhiyong},
	title = {{DVC}: An End-to-End Deep Video Compression Framework},
	booktitle = {Proc. IEEE/CVF Conf. Comput. Vis. Pattern Recognit. (CVPR)},
	month = {June},
	year = {2019},
	pages = {10998--11007},
	doi = {10.1109/CVPR.2019.01126}}

@ARTICLE{Video,
	author={Wang, Sixian and Dai, Jincheng and Liang, Zijian and Niu, Kai and Si, Zhongwei and Dong, Chao and Qin, Xiaoqi and Zhang, Ping},
	journal={IEEE J. Sel. Areas Commun.}, 
	title={Wireless Deep Video Semantic Transmission}, 
	year={2023},
	volume={41},
	number={1},
	pages={214--229},
	doi={10.1109/JSAC.2022.3221977}}
	%	
	%	\footnote{This work considers a single legitimate receiver (Bob) and a single eavesdropper (Eve). The proposed framework can be extended to scenarios involving multiple receivers and eavesdroppers by carefully balancing trade-offs between communication reliability across legitimate receivers, as well as between security levels against eavesdroppers.}
\end{document}